\documentclass{aa}

\usepackage{graphicx}

\usepackage{txfonts}
\usepackage{bbm}
\usepackage{url}
\usepackage{natbib}
\usepackage{amsmath}

\usepackage[breaklinks=true]{hyperref}
\hypersetup{colorlinks=true,urlcolor=blue,citecolor=blue,pdfborder= 0 0 0}
\bibpunct{(}{)}{;}{a}{}{,}
\usepackage{xcolor}
\DeclareMathOperator*{\argmax}{arg\,max}
\DeclareMathOperator*{\argmin}{arg\,min}
\DeclareMathOperator*{\med}{med}

\begin{document}

   \title{TOPz: Photometric redshifts using template fitting applied to the GAMA survey}

   \author{E.~Tempel\inst{1,2}
        \and
        J.~Laur\inst{1}
        \and
        Z.~R.~Jones\inst{1}
        \and
        R.~Kipper\inst{1}
        \and
        L.~J.~Liivam\"agi\inst{1}
        \and
        D.~Pandey\inst{1}
        \and
        G.~Sakteos\inst{1}
        \and
        A.~Tamm\inst{1}
        \and
        A.~N.~Triantafyllaki\inst{1}
        \and
        T.~Tuvikene\inst{1}
          }

   \institute{Tartu Observatory, University of Tartu, Observatooriumi 1, 61602 Tõravere, Estonia\\
            \email{elmo.tempel@ut.ee}
            \and
            Estonian Academy of Sciences, Kohtu 6, 10130 Tallinn, Estonia
            }

    \abstract
   {Accurate photometric redshift (photo-z) estimation is crucial for cosmological and galaxy evolution studies, especially with the advent of large-scale photometric surveys.}
   {We developed a photo-z estimation code called TOPz (Tartu Observatory Photo-z) and applied it to the GAMA photometric catalogue. Using nine-band photometric data from the GAMA project, we assessed the accuracy of TOPz by comparing its photo-z estimates to available spectroscopic redshifts from GAMA and DESI DR1. The latter extends to $z<2$ and $m_Z<24$, allowing the photo-z accuracy to be validated beyond the GAMA limits.
   }
   {TOPz employs a Bayesian template-fitting approach to estimate photo-z from marginalised redshift posteriors. 
   We generated synthetic galaxy spectra using the CIGALE software and ran template set optimisation. 
   We improved the photometry by applying flux and flux uncertainty corrections.
   An analytical prior was then imposed on the resulting posteriors to refine the redshift estimates.
   }
   {The photo-z estimates produced by TOPz show good agreement with the spectroscopic redshifts in the low-redshift regime ($z<0.5$) where the majority (95\%) of the GAMA spectroscopic redshifts are. 
   We demonstrate the redshift accuracy across various magnitude bins and tested how the flux corrections and posteriors reflect the actual uncertainty of the estimates. For the GAMA sample, the $\sigma_\mathrm{NMAD}=0.012$ for $m_Z<18$ and increases to $\sigma_\mathrm{NMAD}=0.021$ for $m_Z>19$. The outlier fraction ($|\Delta z|/(1+z)>0.1$) in the same magnitude bins increases from 1\% to 5\%.
   We show that the TOPz results are consistent with those obtained from other photo-z codes (EAZY and SFM) applied to the same data set. 
   Additionally, TOPz estimates stellar masses as a by-product, comparable to those calculated by other methods. 
   We have made the full GAMA photo-z catalogue and all the codes and scripts used for the analysis and figures publicly available.
   }
   {TOPz is an advanced photo-z estimation code that integrates flux corrections, physical priors, and template set optimisation to provide state-of-the-art photo-z among competing template-based redshift estimators.
   Future work will focus on incorporating additional photometric data and applying the TOPz algorithm to the J-PAS narrow-band survey, further validating and enhancing its capabilities.
   }


   \maketitle

\section{Introduction}
\label{sec:intro}

Photometric redshifts (photo-z) are essential in cosmology and galaxy evolution studies, allowing the expansion of galaxy samples limited by spectroscopic surveys. New surveys such as the Rubin Observatory, Euclid, and Roman Space Telescope will provide deep photometry over large sky areas that allow the inclusion of faint galaxies in the analysis. Estimating accurate photo-z for them is crucial since the number of objects is too numerous for spectroscopic follow-up. \citet{2019NatAs...3..212S} and \citet{2022ARA&A..60..363N} provide an excellent overview of how current photo-z codes have to be improved to take advantage of the next-generation photometric surveys.

A wide variety of methods have been developed for estimating photo-z. They are divided into two primary categories based on machine learning or template fitting. Machine-learning methods include neural networks \citep{2016PASP..128j4502S, 2018A&A...616A..69B, 2018A&A...609A.111D, 2019A&A...621A..26P, 2021A&A...651A..55S, 2024ApJ...964..130J, 2024MNRAS.527..651T, 2025JCAP...01..097P}, random forests or decision trees \citep{2010ApJ...715..823G, 2013MNRAS.432.1483C}, Gaussian processes \citep{2016MNRAS.462..726A, 2016MNRAS.455.2387A, 2021MNRAS.503.4118S}, a genetic algorithm \citep{2015MNRAS.449.2040H}, or other machine-learning approaches \citep{2020MNRAS.498.5498H, 2021A&A...650A..90A, 2021MNRAS.507.5034R, 2022MNRAS.512.3662D}. 

In the machine-learning approach, there is a need for a spectroscopic training sample, which is also used in methods using direct neighbourhood fitting \citep{2016MNRAS.459.3078D} or scaled flux matching \citep{2021MNRAS.500.1557B}. While these methods provide excellent results, they are limited to the set of galaxy types and redshift ranges covered by the spectroscopic training set \citep[e.g.][]{2008A&A...480..703H, 2017MNRAS.468.4323B, 2024A&A...686A..38T}. Solutions to overcome this are proposed by augmenting the spectroscopic samples with simulated data \citep{2024ApJ...967L...6M} or aiding the photo-z estimation with a classification algorithm \citep{2018A&A...619A..14F}.

Template-fitting-based photo-z estimation overcomes many problems related to methods that use spectroscopic samples as a training set. Template fitting in its general form allows a Bayesian inference \citep{Laplace1774} to estimate photo-z posteriors. However, not all template-fitting algorithms are Bayesian in nature. Additionally, when input templates are physical, template-fitting methods can provide a rough estimate for the physical properties of galaxies. Examples of template-fitting application include the following: EAZY \citep{2008ApJ...686.1503B}, LePhare \citep{1999MNRAS.310..540A, 2006A&A...457..841I}, BPZ \citep{2000ApJ...536..571B}, ZEBRA \citep{2006MNRAS.372..565F}, HyperZ \citep{2000A&A...363..476B}, and TOPz \citep{2022A&A...668A...8L}.  Lately, hybrid methods that combine machine learning and template fitting are emerging \citep{2022MNRAS.513.3719H, 2024MNRAS.530.2012T} or hybrid techniques that combine spectroscopic training sets with template fitting \citep{2016MNRAS.460.1371B} that improve the photo-z accuracy \citep{2017MNRAS.466.2039C}.

Many factors influence the accuracy and performance of different template-fitting photo-z codes and methods. The dependence of the used set of templates has been studied in \citet{2000A&A...363..476B} or the combination of different sets of templates \citep{2018MNRAS.473.2655D}, both aiming to improve the accuracy of the photo-z estimation. \citet{2017A&A...608A...3B} demonstrate that template-fitting photo-z codes are reliable down to the 30th magnitude utilising the MUSE observations \citep{2017A&A...608A...1B}.
The need to constantly improve photo-z estimation methods has been a result of applying it to numerous photometric surveys in the past, such as the PAU survey \citep{2024MNRAS.534.1504N}, KiDS survey \citep{2018A&A...616A..69B}, ALHAMBRA survey \citep{2014MNRAS.441.2891M}, DESI Legacy Imaging Survey \citep{2019ApJS..242....8Z}, miniJPAS survey \citep{2021A&A...654A.101H}, or covering the whole extra-Galactic sky \citep{2016ApJS..225....5B}. \citet{2016MNRAS.457.4005W} point out that photo-z methods tend to be overconfident and underestimate the actual uncertainty when photo-z is compared with spectroscopic measurements, resulting in many occasions of overly optimistic estimates \citep{2019MNRAS.489..663B}.

Comparisons between different photo-z codes have been made in various contexts. They all have allowed the conclusion to be drawn that there is no particular method or set of spectral templates that significantly outperforms others and further work is required to identify how to best select between machine-learning and template-fitting approaches for each individual galaxy \citep{2011MNRAS.417.1891A, 2018PASJ...70S...9T, 2020A&A...644A..31E, 2020MNRAS.499.1587S}. \citet{2013ApJ...775...93D} showed that methods that adopted emission lines in spectral templates and added a zero-point offset to observed fluxes performed better. The best method or combination of techniques is goal-dependent and requires science-specific performance metrics. Further improvements in photo-z can be achieved by constraining the photo-z posterior with the underlying cosmic web from spectroscopic surveys \citep{2020A&A...636A..90S, 2023A&A...672A.150T}. Validation techniques, such as cross-correlation \citep{2018MNRAS.477.1664G, 2023A&A...670A.149N} and methods based on normalising flows \citep{2024AJ....168...80C}, have also been developed. The full photo-z posterior is beneficial for extracting complete information from photo-z codes \citep{2017A&A...599A..62L, 2024A&C....4900886T}. Currently, there is no preferred code for photo-z estimation; this gives space to improve the current codes available or even create new ones.

In our previous paper \citet{2022A&A...668A...8L}, we used a template-fitting approach to photo-z estimation where the physically motivated templates were generated using the CIGALE software \citep{2019A&A...622A.103B, 2022ApJ...927..192Y}. With the upcoming photometric information from surveys utilising narrow-band filters, template fitting is expected to achieve an even higher accuracy than what is described in \citet{2022A&A...668A...8L}, especially with the upcoming J-PAS survey \citep{2021A&A...653A..31B}, which provides photometry in more than 50 filters. The currently available narrow-band photometric surveys with fewer filters, such as COMBO-17 \citep{2004A&A...421..913W}, COSMOS \citep{2009ApJ...690.1250S}, or ALHAMBRA \citep{2014MNRAS.441.2891M}, have shown how photo-z estimation benefits from the narrow-band filters. \citet{2000ApJ...536..571B} demonstrated the impact of the width of the band, the number of bands, and the spacing between bands while computing photo-z with template fitting.

In this paper, we present the improved version of the TOPz (Tartu Observatory Photo-z) code, where we specifically improved the observed flux corrections (zero-point offsets) and template set selection and applied physically motivated priors. We extended its application to the GAMA (Galaxy And Mass Assembly) sample of galaxies. We aim to demonstrate that this approach will be valuable for future surveys, such as 4MOST WAVES \citep{2019Msngr.175...46D}, which is based on the same photometry as the GAMA survey.

The outline of the paper is the following: In Sect.~\ref{sec:topz}, we describe the Bayesian photo-z estimation method. The GAMA sample is described in Sect.~\ref{sec:gama}. In Sect.~\ref{sec:fluxcorr}, we describe the scheme for correcting observational flux and flux uncertainties for photo-z estimation. In Sect.~\ref{sec:templates}, we describe the construction and selection of the templates for the TOPz photo-z code and in Sect.~\ref{sec:stargal} we describe how the TOPz can be used for star-galaxy identification. In Sect.~\ref{sec:prior}, we describe the construction of physical priors based on the survey volume and luminosity function of galaxies. In Sect.~\ref{sec:topz_gama}, we present the TOPz photo-z results applied to the GAMA data, and in Sect.~\ref{sec:conclusions} we conclude the paper and discuss future directions. The TOPz code has been made publicly available and the computed photo-z for the GAMA data are available in the CDS (see Appendix~\ref{app:gama}). In Appendix~\ref{app:desi} we validate our photo-z estimates using DESI DR1 spectroscopic data beyond the GAMA spectroscopic limits.

\section{TOPz: Bayesian photometric redshifts using template fitting}
\label{sec:topz}

Template-fitting methods involve matching the observed photometric data of galaxies to a set of pre-defined spectral templates, which represent different types of galaxies at various redshifts. The goal is to compute the likelihood that a given template at a particular redshift matches the observed data, allowing us to derive the probability distribution function (PDF) of the galaxy's photo-z.

Within the framework of Bayesian probability, the task of finding redshift $\zeta = \ln(1+z)$\footnote{Following \citet{2018arXiv181205135B} we use $\zeta = \ln(1+z)$ instead of the redshift~$z$.} given the galaxy observed fluxes $F$, flux uncertainties\footnote{In all equations, the notation $F$ includes both, fluxes and flux uncertainties. When necessary, flux and flux uncertainty in a given passband $\alpha$ are designated as $F_\alpha$ and $\sigma_{F_\alpha}$.} and total magnitude $m_0$ is to find the conditional probability $p(\zeta\,|\,F,m_0)$. To simplify the notations, the observed galaxy properties $F$ and $m_0$ for a single galaxy are denoted as $G$ in some equations below, hence $p(\zeta\,|\,F,m_0) \equiv p(\zeta\,|\,G)$. The photo-z probability $p(\zeta\,|\,F,m_0)$ for a single galaxy using template fitting can be expressed as \citep{2000ApJ...536..571B, 2022A&A...668A...8L}
\begin{equation}
    p(\zeta\,|\,F,m_0) =\! \sum\limits_{T\in \mathbf{T}} p(\zeta,T\,|\,F,m_0) \propto\! \sum\limits_{T\in \mathbf{T}} p(\zeta,T\,|\,m_0) p(F\,|\,\zeta,T),
    \label{eq:marginalisation}
\end{equation}
where $p(F\,|\,\zeta,T)$ gives the likelihood that observed galaxy fluxes $F$ correspond to the template $T$ at redshift $\zeta$ (see Sect.~\ref{sec:zlikelihood}) and $p(\zeta,T\,|\,m_0)$ gives the prior information that galaxy with a reference magnitude $m_0$ has a specific spectral template $T$ and redshift $\zeta$ (see Sect.~\ref{sec:prior}).

In the photo-z probability estimation, we use fluxes $F$ to estimate the redshift likelihood $p(F\,|\,\zeta,T)$ and we use total magnitude $m_0$ for the prior $p(\zeta,T\,|\,m_0)$. Magnitude $m_0$ in a reference passband represents the total magnitude of a galaxy, which may differ from the fluxes $F$ utilised in template fitting. For template fitting, galaxy colours (flux ratios) must be accurate. For this reason, the galaxy fluxes for template fitting are usually measured using aperture-corrected fluxes and the same sky mask in all observed bands. Hence, the observed fluxes that provide accurate colours are underestimating the total flux of a galaxy.

For the template fitting, we have to define a set of templates $\mathbf{T}$ that covers all different types of galaxies. The generation of galaxy templates and subsequent template selection for photo-z estimation are described in Sect.~\ref{sec:templates}.

\subsection{Redshift likelihood estimation}
\label{sec:zlikelihood}

\begin{figure}
\centering
    \includegraphics[width=\columnwidth]{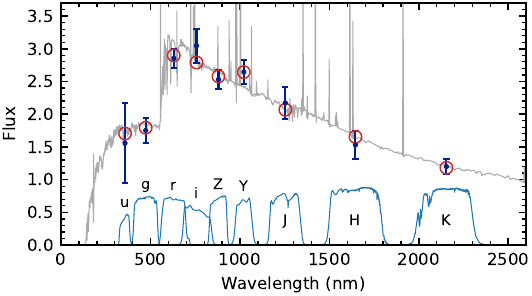}
    \caption{Illustration of the template-fitting approach. The model spectrum is shown with a grey line, and red circles denote the model fluxes in different passbands. The passband transmission curves used for the GAMA data are shown as blue lines. The blue points with error bars are observed fluxes with 2-$\sigma$ errors for a given galaxy. The template spectrum is normalised to best match the observed fluxes.}
    \label{fig:passbands}
\end{figure}

The redshift likelihood for galaxies can be expressed as \citep[see][]{2000ApJ...536..571B}
\begin{equation}
    p(F\,|\,\zeta,T) \propto \frac{1}{\sqrt{F_\mathrm{TT}}} \exp{\left[ -\frac{\chi^2(\zeta,T,a_m)}{2}\right]},
    \label{eq:zlikelihood}
\end{equation}
where $F^{-1/2}_\mathrm{TT}$ is a normalisation parameter that comes from integrating over a nuisance parameter $a$ (see Eq.~\eqref{eq:chi2} and~\eqref{eq:FTT}). As is common in template fitting, the $\chi^2$ is defined as
\begin{equation}
    \chi^2(\zeta,T,a) = \sum\limits_{\alpha} \frac{\left( F_\alpha - aF_{T\alpha} \right)^2}{\sigma^2_{F_{\alpha}}},
    \label{eq:chi2}
\end{equation}
where summation is over all observed fluxes, $F_\alpha$ and $\sigma_{F_{\alpha}}$ are observed flux and flux uncertainty in passband $\alpha$, and $F_{T\alpha}$ is model flux in passband $\alpha$ for template $T$. The parameter $a$ is a nuisance parameter that minimises the $\chi^2(\zeta,T,a)$ for a given template $T$ and redshift $\zeta$ pair. Eq.~\eqref{eq:chi2} is minimised when $a = a_m = F_{\mathrm{OT}} / F_{\mathrm{TT}}$.
The $\chi^2$ when $a=a_m$ is given as
\begin{equation}
    \chi^2(\zeta,T,a_m) = F_{\mathrm{OO}} - \frac{F^2_{\mathrm{OT}}}{F_{\mathrm{TT}}} .
\end{equation}
In the last equation, the quantities $F_\mathrm{XX}$ are defined as
\begin{equation}
    F_\mathrm{OO}=\sum\limits_{\alpha}\frac{F^2_\alpha}{\sigma^2_{F_{\alpha}}}, \quad
    F_\mathrm{TT}=\sum\limits_{\alpha}\frac{F^2_{T\alpha}}{\sigma^2_{F_{\alpha}}}, \quad
    F_\mathrm{OT}=\sum\limits_{\alpha}\frac{F_{\alpha}F_{T\alpha}}{\sigma^2_{F_{\alpha}}} .
    \label{eq:FTT}
\end{equation}

The redshift likelihood normalisation parameter $F^{-1/2}_\mathrm{TT}$ in Eq.~\eqref{eq:zlikelihood} depends on the internal normalisation of the template. For a single template $T$, the template normalisation is irrelevant. However, when combining redshift likelihoods from multiple templates, the internal normalisation of each template affects the redshift likelihood $p(F\,|\,\zeta,T)$. To overcome this problem, we renormalise each template $T$ to achieve the minimum $\chi^2$ when $a_m = 1$. Under this assumption, for arbitrarily normalised templates, the redshift likelihood can be written as
\begin{equation}
    p(F\,|\,\zeta,T) \propto \frac{\sqrt{F_\mathrm{TT}}}{F_\mathrm{OT}} \exp{\left[ -\frac{\chi^2(\zeta,T,a_m)}{2}\right]}.
    \label{eq:zlhood}
\end{equation}
In practical application, Eq.~\eqref{eq:zlhood} is used to compute the redshift likelihood for a given template $T$ at redshift $\zeta$.

Model flux $F_{T\alpha}$ in passband $\alpha$ is computed as
\begin{equation}
    F_{T\alpha} = \frac{\int f_{T,\zeta}(\lambda)W_\alpha(\lambda) \,\mathrm{d}\lambda}
    {\int W_\alpha(\lambda)\,\mathrm{d}\lambda},
    \label{eq:model_flux}
\end{equation}
where $f_{T,\zeta}(\lambda)$ is the specific flux of the template spectra $T$ at redshift $\zeta$ and $W_\alpha$ is window function or transmission curve for passband $\alpha$. Given a model template spectra $f_T(\lambda)$ at redshift zero, the redshifted spectrum $f_{T,\zeta}(\lambda)$ is computed as
\begin{equation}
    f_{T,\zeta}(\lambda) \propto f_T(\lambda\,\mathrm{e}^{-\zeta}).
\end{equation}

Figure~\ref{fig:passbands} illustrates the basics of the template-fitting approach. We compute model fluxes from the model template spectrum in different passbands using Eq.~\eqref{eq:model_flux}. The model fluxes are compared against observed fluxes while including the observed flux uncertainties, and the likelihood that a given model template is good is estimated using Eq.~\eqref{eq:zlhood}.

\begin{figure}
\centering
    \includegraphics[width=\columnwidth]{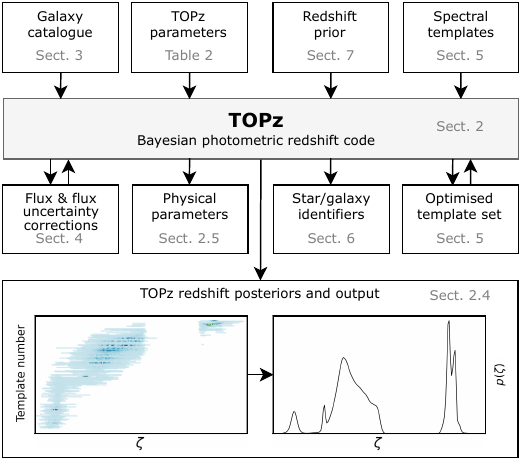}
    \caption{ Schematic overview of the TOPz code, illustrating key inputs and outputs. The two-way arrows indicate that an iterative approach can be applied for further optimisation. Refer to the sections indicated in the corresponding boxes for additional information. See Fig.~\ref{fig:topz_posterior} for the detailed description of the TOPz posteriors and output.} 
    \label{fig:topz}
\end{figure}

\subsection{TOPz: Photometric redshift code}

The Bayesian photo-z estimation based on template fitting is a general approach implemented in many codes. Although all the codes share the basics of the template-fitting approach, the codes do not provide identical photo-z estimates. The reason is the practical implementation and technical details of the template-fitting approach. To have complete control over implemented features and to fully understand the technical details, we developed the TOPz code (Tartu Observatory Photo-z). The TOPz code is a Bayesian photo-z code based on template fitting, incorporating many aspects of the previously developed template-fitting codes. The precursor of the current implementation of the code is given in \citet{2022A&A...668A...8L}, but many of the details in the current TOPz implementation are significantly improved upon.

To achieve the best possible photo-z using the template-fitting approach, we have to pay attention to the following aspects: galaxy template spectra generation and subsequent template set optimisation (see Sect.~\ref{sec:templates}), observed flux and flux uncertainty corrections (see Sect.~\ref{sec:fluxcorr}), physically motivated priors for photo-z (see Sect.~\ref{sec:prior}). All these aspects are equally important and affect the accuracy and precision of the estimated photo-z. In Fig.~\ref{fig:topz}, we show the schematic overview of the TOPz code, highlighting the key inputs and outputs.

The TOPz code is written in Fortran programming language and is publicly available. Together with the Fortran code, we make available all Python scripts used to run the TOPz code and generate all the figures presented in this paper.

\subsection{TOPz: Marginalised redshift posterior}

The main output of the TOPz code is a two-dimensional probability distribution $p(\zeta,T\,|\,F,m_0)$ for each input source given the observed fluxes, flux uncertainties and total magnitude. This probability distribution is a combination of redshift likelihood and prior information. An example of a two-dimensional probability distribution for a rare source with many maxima is shown in Fig.~\ref{fig:topz_posterior}. Most of the sources have only one dominant maximum (see Sect.~\ref{sec:photoz_gama} for the statistics in the GAMA sample).

For redshift estimation, the two-dimensional probability $p(\zeta,T\,|\,F,m_0)$ is marginalised over all input templates (see Eq.~\eqref{eq:marginalisation}), which provides us a redshift posterior distribution $p(\zeta\,|\,F,m_0)$. An example of a redshift posterior for a single galaxy is shown in the lower panel in Fig.~\ref{fig:topz_posterior}. For the majority of galaxies, there is only one dominant redshift peak. However, for many fainter galaxies the redshift posterior has more than one peak. To simplify the notations below, the one-dimensional redshift posterior distribution for galaxy $G$ (with parameters $F$ and $m_0$) given an input template set $\mathbf{T}$ is designated as $p^G_\mathbf{T}(\zeta)$.

It is known that the width of the peaks in the photo-z posterior usually underestimates the true uncertainty of the redshift estimate. To mitigate this problem, the marginalised redshift distribution can be convolved with a kernel that smooths the redshift posterior distribution and increases the width of the individual redshift peaks. In TOPz code, we have the option to convolve the redshift posterior $p^G_\mathbf{T}(\zeta)$ with a Gaussian kernel $K(\zeta ; m_0)$, where the width of the kernel depends on the galaxy total magnitude $m_0$. The kernel width dependence on galaxy magnitude should be explicitly estimated for a given data set. In all equations given in Sect.~\ref{sect:topz_output}, we assume that the redshift posterior distribution is a convolved distribution.

\subsection{TOPz: Output}
\label{sect:topz_output}

In TOPz code, we identify three dominant peaks in the redshift posterior $p^G_\mathbf{T}(\zeta)$, designated as 'best', 'alt1', and 'alt2'. These peaks in the posterior distribution are defined so they do not overlap in redshift. The peaks are ordered so that
\begin{equation}
    p^G_\mathbf{T}(\zeta^\mathrm{best}_\mathrm{peak}) >
    p^G_\mathbf{T}(\zeta^\mathrm{alt1}_\mathrm{peak}) >
    p^G_\mathbf{T}(\zeta^\mathrm{alt2}_\mathrm{peak}) .
\end{equation}
Alternatively, the peaks can be ordered based on the area below the peaks (see Eq.~\eqref{eq:ppeak}). As follows, we define parameters for the dominant 'best' peak. The parameters for 'alt1' and 'alt2' peaks are defined similarly.

\begin{figure}
\centering
    \includegraphics[width=\columnwidth]{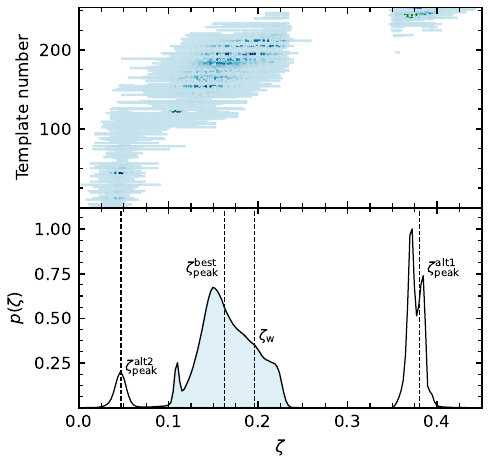}
    \caption{TOPz redshift posterior for a source with many maxima and a spectroscopic redshift of $z \approx 0.21$ ($\zeta \approx 0.19$). The upper panel displays a 2D posterior distribution, with all templates arranged in order of increasing weighted $\zeta$ values. The colour of each point represents the photo-z likelihood of the corresponding template. The lower panel illustrates the marginalised posterior distribution for the same source. Dashed vertical lines indicate the $\zeta$ values for all peaks along with $\zeta_{w}$. For details on the definitions of the three peaks and the probability-weighted $\zeta_{w}$, refer to Sect.~\ref{sect:topz_output}. The shaded area in the lower panel represents the range of the best peak, bounded by the minimum ($\zeta^{\mathrm{best}}_{\min}$) and maximum ($\zeta^{\mathrm{best}}_{\max}$) values.}
    \label{fig:topz_posterior}
\end{figure}

For the best peak $\zeta^\mathrm{best}_\mathrm{peak}$ we define the peak boundaries $\zeta^{\mathrm{best}}_{\min}$ and $\zeta^{\mathrm{best}}_{\max}$ as follows
\begin{align}
     \zeta^{\mathrm{best}}_{\min} = \argmin_{\zeta < \zeta^\mathrm{best}_\mathrm{peak}} \left\{ \zeta^\mathrm{best}_\mathrm{peak}\!-\!\zeta  \,\middle|\, \frac{\mathrm{d}p^G_\mathbf{T}(\zeta)}{\mathrm{d}\zeta} = 0 \,\&\, \frac{p^G_\mathbf{T}(\zeta)}{p^G_\mathbf{T}(\zeta^\mathrm{best}_\mathrm{peak})} < f_\mathrm{peak}  \right\} , \\
     \zeta^{\mathrm{best}}_{\max} = \argmin_{\zeta > \zeta^\mathrm{best}_\mathrm{peak}} \left\{ \zeta\!-\!\zeta^\mathrm{best}_\mathrm{peak}  \,\middle|\, \frac{\mathrm{d}p^G_\mathbf{T}(\zeta)}{\mathrm{d}\zeta} = 0 \,\&\, \frac{p^G_\mathbf{T}(\zeta)}{p^G_\mathbf{T}(\zeta^\mathrm{best}_\mathrm{peak})} < f_\mathrm{peak} \right\} .
\end{align}
The peak boundaries are defined as the first minima before and after the peak maximum that are lower than $f_\mathrm{peak}\cdot p^G_\mathbf{T}(\zeta^\mathrm{best}_\mathrm{peak})$, where $f_\mathrm{peak} < 1.0$ is a free parameter in the TOPz code. It is recommended to set the $f_\mathrm{peak}$ to a rather low value (e.g. 0.005) to capture the majority of the peak and to avoid spurious small minima in the posterior distribution. The main aim of $f_\mathrm{peak}$ is to identify peaks that are clearly separated from each other as shown in Fig.~\ref{fig:topz_posterior}. Setting the $f_\mathrm{peak}=0.0$ identifies the full posterior as one dominant peak. The redshift value, where the posterior peak has a maximum, is designated as $\zeta^\mathrm{best}_\mathrm{peak}$ and defined as
\begin{equation}
    \zeta^\mathrm{best}_\mathrm{peak} = \argmax_{\zeta^{\mathrm{best}}_{\min} < \zeta < \zeta^{\mathrm{best}}_{\max}} p^G_\mathbf{T}(\zeta) .
\end{equation}
To simplify the notations, we do not specifically indicate that $\zeta^\mathrm{best}_\mathrm{peak}$ for galaxy $G$ depends on the input template set $\mathbf{T}$. Where needed, the dependence is explicitly noted as $\zeta^\mathrm{best}_\mathrm{peak}(G,\mathbf{T}) \equiv \zeta^\mathrm{best}_\mathrm{peak}$. The same simplification applies to other peak properties defined in this section.

Additionally, we define the probability-weighted redshift value for each peak
\begin{equation}
    \zeta^{\mathrm{best}}_{\mathrm{w}} = 
    \frac{\int\limits_{\zeta^{\mathrm{best}}_{\min}}^{\zeta^{\mathrm{best}}_{\max}}
    \zeta p^G_\mathbf{T}(\zeta)\mathrm{d}\zeta} 
    {\int\limits_{\zeta^{\mathrm{best}}_{\min}}^{\zeta^{\mathrm{best}}_{\max}}
    p^G_\mathbf{T}(\zeta) \mathrm{d}\zeta} .
\end{equation}
The summed probability in the redshift posterior that is covered by the best peak is defined as
\begin{equation}
    p^\mathrm{best}_\mathrm{peak} = \frac{\int\limits_{\zeta^{\mathrm{best}}_{\min}}^{\zeta^{\mathrm{best}}_{\max}} p^G_\mathbf{T}(\zeta)\mathrm{d}\zeta} {\int\limits_{-\inf}^{\inf} p^G_\mathbf{T}(\zeta)\mathrm{d}\zeta} .
    \label{eq:ppeak}
\end{equation}
We also provide the posterior weighted redshift value over the full redshift range. It is defined as
\begin{equation}
    \zeta_{\mathrm{w}} = 
    \frac{\int\limits_{-\inf}^{-\inf}
    \zeta p^G_\mathbf{T}(\zeta)\mathrm{d}\zeta} 
    {\int\limits_{-\inf}^{-\inf}
    p^G_\mathbf{T}(\zeta) \mathrm{d}\zeta} .
\end{equation}
The difference between $\zeta_{\mathrm{w}}$ and $\zeta^{\mathrm{best}}_{\mathrm{w}}$ together with the $p^\mathrm{best}_\mathrm{peak}$ value can be used to evaluate the reliability of the estimated photo-z. If the difference between $\zeta_{\mathrm{w}}$ and $\zeta^{\mathrm{best}}_{\mathrm{w}}$ is large or the $p^\mathrm{best}_\mathrm{peak}$ value is low, then it indicates that the posterior has more than one peak and the photo-z solution is not unique.

The marginalised redshift posterior $p^G_\mathbf{T}(\zeta)$ distribution provides the uncertainty of the TOPz photo-z estimate. For approximate calculations, we compute the standard deviation of the redshift posterior distribution. The standard deviation is a good measure of the redshift uncertainty when the posterior distribution has only one peak. The computed standard deviation estimates should be treated with care for multi-peaked posteriors. The standard deviation, the photo-z uncertainty estimate for $\zeta_{\mathrm{w}}$, is computed as
\begin{equation}
    \sigma^2_{\zeta_{\mathrm{w}}} = 
    \frac{\int\limits_{-\inf}^{-\inf}
    p^G_\mathbf{T}(\zeta) \left(\zeta-\zeta_{\mathrm{w}}\right)^2 \mathrm{d}\zeta} 
    {\int\limits_{-\inf}^{-\inf}
    p^G_\mathbf{T}(\zeta) \mathrm{d}\zeta} .
    \label{eq:zeta_sigma}
\end{equation}
Photo-z uncertainty estimates for $\zeta^{\mathrm{best}}_{\mathrm{w}}$, $\zeta^{\mathrm{alt1}}_{\mathrm{w}}$, and $\zeta^{\mathrm{alt2}}_{\mathrm{w}}$ are computed similarly while limiting the integration over a given peak.

\subsection{TOPz: Physical parameters ot galaxies}
\label{sec:phys_params}

For template fitting, we use physical model templates (see Sect.~\ref{sec:templates}). Using the physical model templates allows us to estimate a galaxy's physical parameters, such as total stellar mass, star formation rate, etc., to a good approximation, as these integrated galaxy parameters corresponding to each model template are known a priori. 

Below, we describe how we estimate the total stellar mass. All other physical parameters can be estimated similarly and are not explicitly defined here. The total stellar mass $M_\mathrm{star}$ of a galaxy is estimated while marginalising over the two-dimensional photo-z posterior

\begin{equation}
    M_\mathrm{star} = \frac{\sum\limits_{T\in\mathbf{T}}\sum\limits_{\zeta} p(F\,|\,\zeta,T) a_m M^T_\mathrm{star} \left(\frac{D_L(\zeta)}{D_L(\zeta_T)}\right)^2 }{\sum\limits_{T\in\mathbf{T}}\sum\limits_{\zeta} p(F\,|\,\zeta,T)} ,
\end{equation}
where $M^T_\mathrm{star}$ and $\zeta_T$ are model template stellar mass and redshift, respectively. The $D_L(\zeta)$ is a luminosity distance at redshift $\zeta$, and $a_m$ is the model template scaling factor from template fitting. In general, the stellar mass can also be estimated at a fixed redshift (e.g. from spectroscopic observations) while marginalising only over templates $\mathbf{T}$.

In Sect.~\ref{sec:topz_gama}, we show how the TOPz estimated stellar masses compare with the stellar masses provided in the GAMA data (see Sect.~\ref{sec:gama}). We emphasise that the TOPz does not perform any physical modelling of galaxies, it relies on the physical spectral templates. A dedicated software (e.g. CIGALE or ProSpect) should be used to model galaxy spectral energy distribution in detail.

\section{GAMA galaxy sample}
\label{sec:gama}

Galaxy And Mass Assembly (GAMA) survey is a spectroscopic survey carried out on the Anglo-Australian Telescope with the AAOmega multi-object spectrograph \citep{2011MNRAS.413..971D}. GAMA results are released in 'Data Management Units' and value-added catalogues are released using other publicly available independent observations to complement its data. In this paper, we take the photometric input data from the GAMA project DR4 \citep{2022MNRAS.513..439D} gkvScienceCatv02 catalogue, which is a 'science-ready' subset of the entire GAMA input catalogue \citep{2020MNRAS.496.3235B}. We exclude duplicated and masked objects from our analysis.

Photometry over a wide range of wavelengths is incorporated into GAMA from various imaging sources such as GALEX, WISE, Herschel observations, and most notably for this paper in optical and near-infrared bands from the European Southern Observatory (ESO) VST KiDS \citep{2013ExA....35...25D} and ESO VISTA VIKING surveys \citep{2013Msngr.154...32E}. For GAMA DR4, the image data from all the included surveys were uniformly regridded and retiled onto rescaled KiDS tiles. Source detection and determination of fluxes (as well as corresponding errors) in all filters were subsequently done using the ProFound package \citep[see][for details]{2018MNRAS.476.3137R, 2020MNRAS.496.3235B}. The photometry is corrected for foreground Galactic extinction employing Planck maps and potentially contaminating bright stars are masked out using GAIA DR2 data. We use the total fluxes with global sky subtraction ({\it flux\_t} columns in the GAMA database), which is expected to represent the total galaxy flux.

In the current paper, we test TOPz on deep nine-band photometry available in the GAMA survey footprint: u, g, r, i from the KiDS and Z, Y, J, H, Ks from the VIKING imaging surveys. Despite more filters being available, we restrict ourselves to the optical passbands as \citet{2020AJ....159..258G} analysed the impact of near-infrared and near-ultraviolet filters on the photo-z estimation and concluded that deeper optical photometry is more beneficial than adding additional filters. It requires a dedicated study to understand the impact of ultraviolet and infrared filters on the GAMA data and how these filters affect the photo-z performance for different types of galaxies with varying magnitudes and redshifts.

In terms of photometry, the GAMA survey is based on the same photometry as the upcoming WAVES \citep{2019Msngr.175...46D}. The target selection for the WAVES is utilising the photo-z estimates based on the same nine-band photometry. For the WAVES target selection, TOPz is one of the four photo-z methods that are optimally combined (in prep).

GAMA provides a catalogue of photometrically derived stellar mass estimates for galaxies determined through stellar population synthesis, which is taken from the StellarMassesGKVv24 catalogue in the GAMA database for the equatorial and G23 survey regions subsets. The procedure is described in detail in \citet{2011MNRAS.418.1587T} while any updates are listed in \citet{2020MNRAS.496.3235B}. We will use these as a test to compare the stellar masses with those corresponding to the templates selected by our method (see Sect.~\ref{sec:phys_params}).

Since GAMA does not strictly include only galaxies but rather gives sources a type designation such as `galaxy', `star' or `ambiguous' (additional option being `artefact'), we will also test if we can recover this classification with appropriate templates (see Sect.~\ref{sec:stargal}). For classification analysis, we use only objects marked as `galaxy' or `star' in the GAMA database.

\begin{figure}
\centering
    \includegraphics[width=\columnwidth]{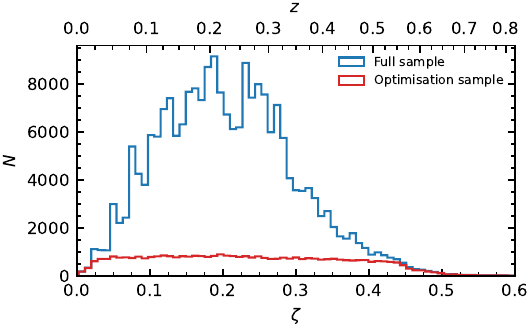}
    \caption{Spectroscopic redshift distribution in the GAMA catalogue (blue line) and a subset of the spectroscopic sample (red line) used for template set optimisation (see Sect.~\ref{sec:templates}).}
    \label{fig:gama_specz_hist}
\end{figure}

\begin{figure}
\centering
    \includegraphics[width=\columnwidth]{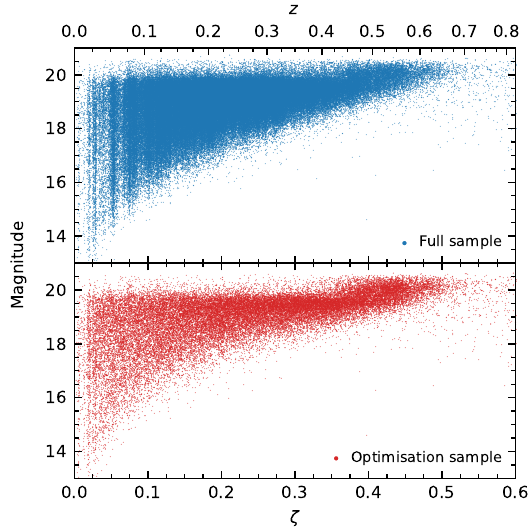}
    \caption{Galaxy redshift $\zeta$ as a function of observed magnitude $m_r$ for the full sample (blue dots) and for the galaxy sample (red dots) used in template set optimisation.}
    \label{fig:gama_specz}
\end{figure}

To construct our spectroscopic input sample, we applied the following conditions. We include galaxies within the spectroscopic redshift range between $0.002 < z < 1.0$ and require the normalised redshift quality {\tt NQ} to be greater than 3. We are further considering only sources within the GAMA spectroscopic survey regions, removing duplicates and objects affected by nearby bright stars. The final input catalogue consists of 221\,108 galaxies. In Figs.~\ref{fig:gama_specz_hist} and \ref{fig:gama_specz}, we show the redshift distribution for the GAMA spectroscopic catalogue used in the current paper. Note that the spectroscopic sample also includes redshifts from external sources (DESI and DEVILS), which go beyond the GAMA spectroscopic redshift limit. However, the spectroscopic sample for the fainter sources is incomplete.

The full photometric GAMA catalogue contains 1\,146\,985 objects and includes all types of sources: `galaxy', `star' or `ambiguous' (specified by the {\tt uberclass} variable). Only objects classified as galaxies with spectroscopic redshifts are used for the photo-z analysis in this paper. Additionally, we will run the TOPz code using the full photometric catalogue and make available the photo-z for all photometric sources in the GAMA database (See Appendix~\ref{app:gama}).

\section{Observed flux and flux uncertainty corrections}
\label{sec:fluxcorr}

\begin{table}
\caption{Flux and flux uncertainty corrections applied to the GAMA data for nine passbands.}
\label{table:fluxcor_param}
\centering
\begin{tabular}{c c c c}
\hline\hline
Passband & $f^\alpha_\mathrm{bias}$ & $f^\alpha_{\sigma,\mathrm{rel}}$ & $f^\alpha_{\sigma,\mathrm{scaling}}$ \\
\hline
   u  & 1.0585 &  0.0500 &  1.3221 \\
   g  & 0.9935 &  0.0000 &  1.3246 \\
   r  & 1.0098 &  0.0000 &  1.0000 \\
   i  & 0.9890 &  0.0000 &  1.0599 \\
   Z  & 0.9566 &  0.0000 &  1.1819 \\
   Y  & 0.9702 &  0.0000 &  1.2295 \\
   J  & 1.0493 &  0.0000 &  1.2112 \\
   H  & 1.0083 &  0.0000 &  1.4387 \\
   Ks & 1.0198 &  0.0000 &  1.6254 \\
\hline
\end{tabular}
\tablefoot{$f^\alpha_\mathrm{bias}$ is linear flux bias correction, $f^\alpha_{\sigma,\mathrm{rel}}$ is relative uncertainty added to the observed flux uncertainty and $f^\alpha_{\sigma,\mathrm{scaling}}$ defines the scaling parameter for flux uncertainty. See Sect.~\ref{sec:fluxcorr} for more details.}
\end{table}

When we fit model templates with observed fluxes, we assume that observed fluxes are accurate (bias-free) and flux errors take into account all statistical and systematic uncertainties. In practice, this assumption is not always valid, and some corrections to fluxes and flux uncertainties might be necessary. In the TOPz code, we have implemented zero-point corrections. First, we test and correct any systematic bias (e.g. photometric zero-point offsets) in observed fluxes. Second, we adjust observed flux uncertainties while adding a potentially missing systematic error component.

For flux and flux uncertainty corrections, we use the full GAMA spectroscopic redshift sample, where we eliminate any systematic outliers. The spectroscopic sample that we use to correct observed fluxes and flux uncertainties satisfies the following criteria:
\begin{itemize}
    \item $\left|\zeta_\mathrm{w}^\mathrm{best} - \zeta_\mathrm{spec}\right| < \mathrm{delta}\_\zeta$, where $\mathrm{delta}\_\zeta = 0.1$;
    \item the reduced $\chi^2$ is smaller than 2.0 for the best-fitting template at spectroscopic redshift.
\end{itemize}
This limits the sample to galaxies, where clear photo-z outliers and bad SED fits are excluded. For individual fluxes in different passbands, we use the additional criteria:
\begin{itemize}
    \item both, flux and flux uncertainty, must exist and observed flux should be flagged as good (e.g. not affected by neighbouring source or imaging artefacts);
    \item computed model fluxes are limited to the best-fitting template at spectroscopic redshift;
    \item clear outliers (bad observed flux estimates) are excluded while requiring $ \left| F_\alpha - a_\mathrm{m}F_{T\alpha} \right| / \sigma_{F_{\alpha}} < \mathrm{delta}\_\mathrm{flux}$\footnote{Notations are the same as in Eq.~\eqref{eq:chi2}, where $\alpha$ indicates a given passband and $a_\mathrm{m}F_{T\alpha}$ is the best-fitting model flux.}, where $\mathrm{delta}\_\mathrm{flux} = 3.0$;
    \item observed flux should have sufficient signal-to-noise ratio, $F_\alpha / \sigma_{F_{\alpha}} > \mathrm{SN}\_\mathrm{limit}$, where $\mathrm{SN}\_\mathrm{limit} = 5.0$.
\end{itemize}
For each galaxy in the sample, the number of used passbands can be different, and for each passband, we can have a different number of galaxies used for flux and flux uncertainty corrections. We designate $\mathbf{S}_\alpha$ as the sample of galaxies for a given passband $\alpha$ that satisfies all the abovementioned criteria. In the GAMA sample, we removed about 10 per cent of objects in all passbands, except the $u$-band, where we removed about one third of the objects.

Observed fluxes are corrected assuming a linear bias correction $f^\alpha_\mathrm{bias}$ and the corrected flux $F^\prime_\alpha$ in passband $\alpha$ is given as
\begin{equation}
    F^\prime_\alpha = f^\alpha_\mathrm{bias} F_\alpha.
    \label{eq:fbiasflux}
\end{equation}
Assuming that the flux bias correction originates from photometric zero-point calibration, we apply the same bias correction to observed flux uncertainties, where the corrected flux error $\sigma^\prime_{F_{\alpha}}$ is given as
\begin{equation}
    \sigma^\prime_{F_{\alpha}} = f^\alpha_\mathrm{bias} \sigma_{F_{\alpha}}.
\end{equation}
The flux bias $f^\alpha_\mathrm{bias}$ is estimated as
\begin{equation}
    f^\alpha_\mathrm{bias} = \med \left\{ f^\alpha_{i}, \, i \in \mathbf{S}_\alpha \right\},
    \label{eq:fbias}
\end{equation}
where $f^\alpha_{i}$ is the ratio of best-fitting model flux over observed flux in a given passband $\alpha$
\begin{equation}
    f^\alpha_{i} = \frac{a^{i}_\mathrm{m}F^{i}_{T\alpha}}{F^{i}_\alpha}.
    \label{eq:falphai}
\end{equation}
In Eq.~\eqref{eq:fbias}, the median is taken over all galaxies and fluxes that satisfy the abovementioned criteria.

Observed flux uncertainties are corrected using the following equation:
\begin{equation}
    \left(\sigma^\mathrm{cor}_{F_{\alpha}}\right)^2 = \left(f^\alpha_{\sigma,\mathrm{scaling}}\right)^2 \left[  \left( \sigma^\prime_{F_{\alpha}} \right)^2 +
    \left( f^\alpha_{\sigma,\mathrm{rel}}F^\prime_\alpha \right)^2 \right] + \sigma^2_\mathrm{model}(\alpha,G).
    \label{eq:sigmacor}
\end{equation}
In the last equation, $f^\alpha_{\sigma,\mathrm{rel}}$ defines the additional relative flux uncertainty added to the observed flux uncertainty in quadrature. The factor $f^\alpha_{\sigma,\mathrm{scaling}}\geqslant 1.0$ scales the observed flux uncertainties so that the mean reduced $\chi^2$ value is close to 1.0. Both factors, $f^\alpha_{\sigma,\mathrm{rel}}$ and $f^\alpha_{\sigma,\mathrm{scaling}}$, are computed using the full sample for photometric corrections and are the same for all galaxies. These factors depend only on the passband. The last factor $\sigma^2_\mathrm{model}(\alpha,G)$ in Eq.~\eqref{eq:sigmacor} adds additional model flux uncertainty that is computed individually for each galaxy. Below, we describe how these factors are computed. For template fitting in Eq.~\eqref{eq:chi2} we use $F^\mathrm{cor}_\alpha\equiv F^\prime_\alpha$ and $\sigma^\mathrm{cor}_{F_{\alpha}}$ instead of $F_\alpha$ and $\sigma_{F_{\alpha}}$.

The additional relative error $f^\alpha_{\sigma,\mathrm{rel}}$ is computed assuming that the scaled difference between observed and model flux is independent of the observed flux. In mathematical terms, the parameter $f^\alpha_{\sigma,\mathrm{rel}}$ is determined so that the linear regression between $\log(F^\prime_\alpha)$ and $\log(\Delta F_\sigma)$ has zero slope, where
\begin{equation}
    \Delta F_\sigma^2 = \frac{\left(F^\prime_\alpha - a_\mathrm{m}F_{T\alpha}\right)^2}{\left(\sigma^\prime_{F_{\alpha}}\right)^2 + \left( f^\alpha_{\sigma,\mathrm{rel}}F^\prime_\alpha \right)^2}.
\end{equation}
When observed fluxes are bias-free and flux uncertainties include all relevant statistical and systematic error components, then the required additional relative error $f^\alpha_{\sigma,\mathrm{rel}}\approx 0$.

The uncertainty scaling factor $f^\alpha_{\sigma,\mathrm{scaling}}$ is estimated as
\begin{equation}
    \left(f^\alpha_{\sigma,\mathrm{scaling}}\right)^2 = 
    \max\left\{ 1.0, \frac{1}{\left|\mathbf{S}_\alpha\right|} \sum\limits_{\mathbf{S}_\alpha} \Delta F_\sigma^2 \right\} ,
    \label{eq:fluxscaling}
\end{equation}
where $\left|\mathbf{S}_\alpha\right|$ is the number of galaxies in a sample $\mathbf{S}_\alpha$. We only scale the observed fluxes upwards when the flux uncertainties underestimate the statistical difference between sigma-scaled observed and model fluxes.

To account for the variance in template fitting and potentially insufficient coverage of model templates, we defined an additional model template dependent systematic uncertainty $\sigma^2_\mathrm{model}(\alpha,G)$ as
\begin{equation}
    \sigma^2_\mathrm{model}(\alpha,G) = \min\left\{ f_\mathrm{mod}^\mathrm{lim}{\sigma^{\prime \,2}_{F_{\alpha}}} , \sigma^{\mathrm{mod}\,2}_{F_{T\alpha}} \right\} + \left( f_\mathrm{mod}^\mathrm{diff} \left|F^\mathrm{mod}_{T\alpha} - F_\alpha\right| \right)^2,
\end{equation}
where $f_\mathrm{mod}^\mathrm{lim}$ and $f_\mathrm{mod}^\mathrm{diff}$ are user defined scaling factors. Model flux $F^\mathrm{mod}_{T\alpha}$ and model flux uncertainty $\sigma^{\mathrm{mod}\,2}_{F_{T\alpha}}$ are estimated as weighted average over all templates
\begin{equation}
    F^\mathrm{mod}_{T\alpha} = \frac{\sum\limits_{T\in\mathbf{T}}\sum\limits_{\zeta} p(F\,|\,\zeta,T) a_m F_{T\alpha}}{\sum\limits_{T\in\mathbf{T}}\sum\limits_{\zeta} p(F\,|\,\zeta,T)} ,
\end{equation}
\begin{equation}
    \sigma^{\mathrm{mod}\,2}_{F_{T\alpha}} = \frac{\sum\limits_{T\in\mathbf{T}}\sum\limits_{\zeta} p(F\,|\,\zeta,T) \left(a_m F_{T\alpha}-F^\mathrm{mod}_{T\alpha}\right)^2}{\sum\limits_{T\in\mathbf{T}}\sum\limits_{\zeta} p(F\,|\,\zeta,T)} ,
\end{equation}
where the summation is over all model templates $\mathbf{T}$ and redshifts. Weights $p$ are proportional to the model likelihoods given with Eq.~\eqref{eq:zlhood}.

As explained above, the corrections are computed one passband at a time. In principle, a correction in one passband affects the corrections in another since the best-fitting template might change after corrections. To account for this, we compute the flux and flux-uncertainty corrections iteratively, including the corrections from previous iterations. The corrections converge very quickly, and we used five iterations for the GAMA sample. The same approach was used in \citet{2022A&A...668A...8L}. This iterative process takes into account the flux scaling over the entire SED. To compute the flux and flux uncertainty corrections, we need a set of model templates $\mathbf{T}$, which are generated as described in Sect.~\ref{sec:templates}. The first set of templates is generated using observed fluxes and uncertainties. When we computed the flux and flux uncertainty corrections, we generated new templates using corrected values. In practice, the corrections change very little, and there was no need to run the template generation process iteratively.

\begin{figure}
\centering
    \includegraphics[width=\columnwidth]{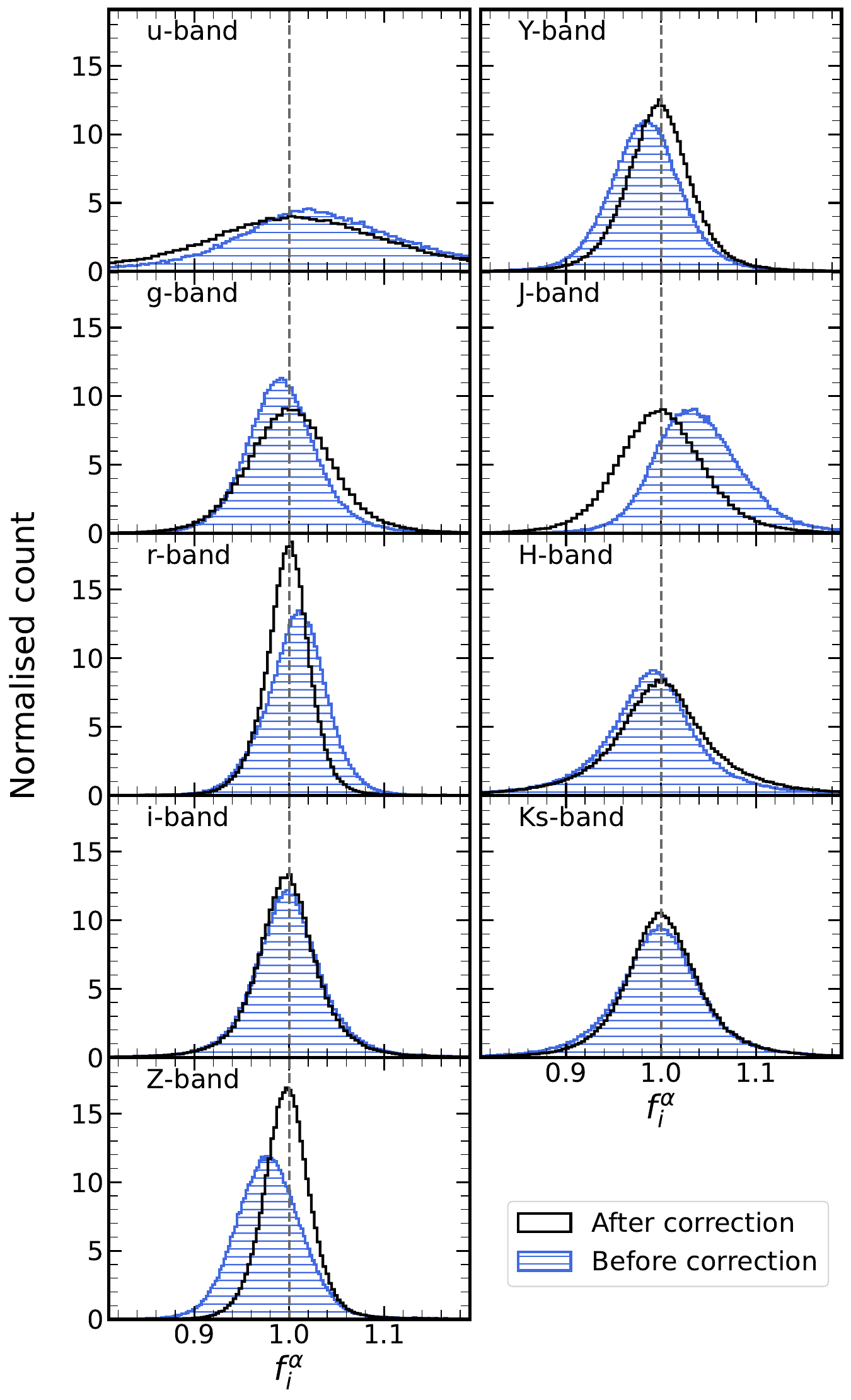}
    \caption{Flux bias corrections for the GAMA spectroscopic sample. The distribution of flux ratios $f^\alpha_i$ (model flux divided by observed flux) for the best-fitting spectral template at spectroscopic redshift. Each panel shows the $f^\alpha_i$ distribution for different passbands. The area under the histograms is normalised to unity. The blue-shaded region shows the flux bias before any corrections. The black line shows the flux bias $f^\alpha_i$ distribution after corrections. The flux bias correction is defined with Eqs.~\eqref{eq:fbiasflux}--\eqref{eq:fbias}. See Sect.~\ref{sec:fluxcorr} for a detailed description of the flux and flux uncertainty corrections. As expected, the flux bias correction shifts the distribution median closer to unity (dashed vertical line). Table~\ref{table:fluxcor_param} shows the flux bias correction values for each passband.} 
    \label{fig:fluxcorbias}
\end{figure}

\begin{figure}
\centering
\includegraphics[width=\columnwidth]{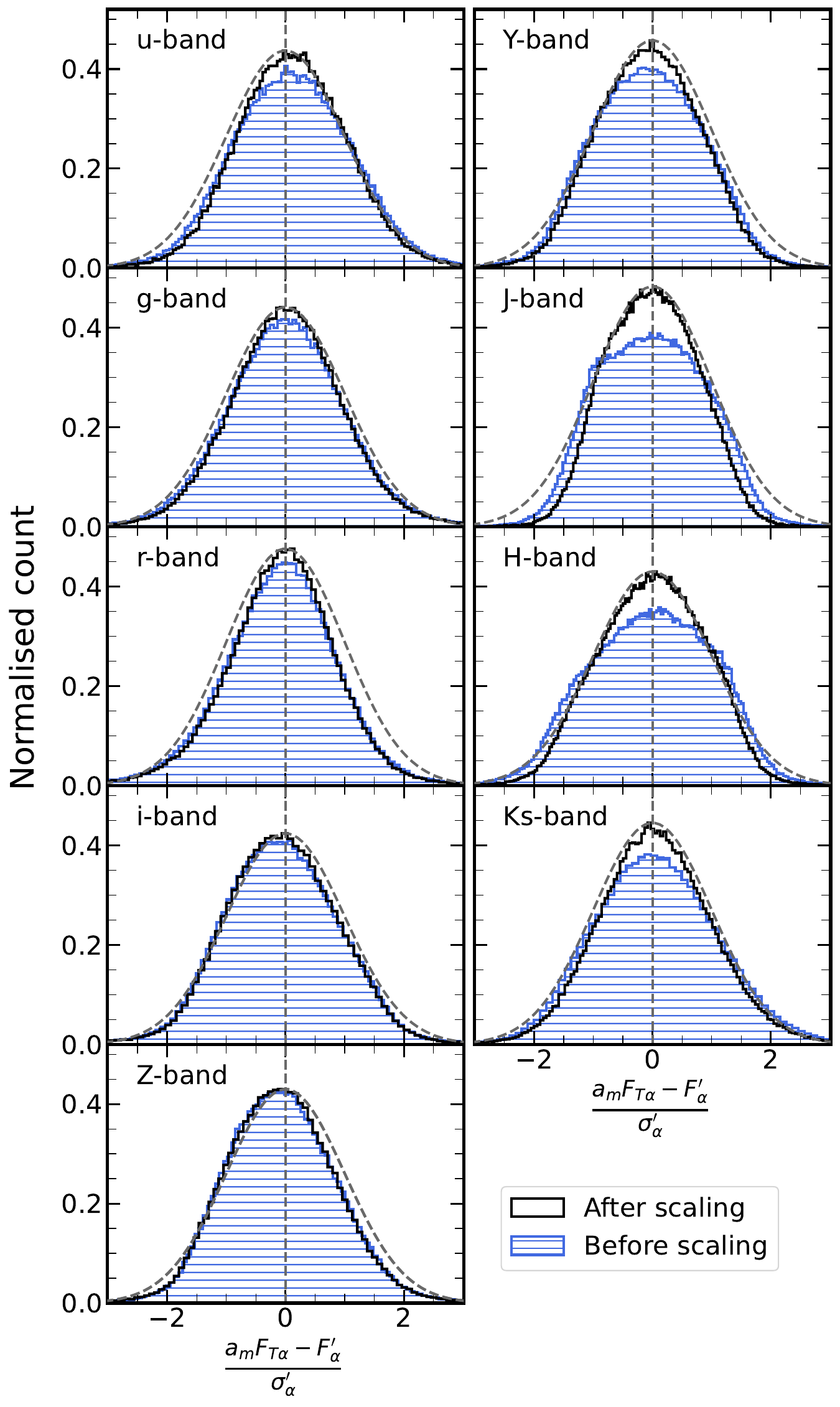}
    \caption{Difference between the model and observed flux divided by the observed flux uncertainty for the best-fitting spectral template at spectroscopic redshift. Each panel shows the distribution for different passbands. The blue-shaded histogram shows the distribution before flux uncertainty scaling. The solid black histogram shows the distribution after flux uncertainty scaling. The dashed grey line shows the Gaussian distribution with unit variance. See Sect.~\ref{sec:fluxcorr} for a detailed description of the flux and flux uncertainty corrections. As expected, the flux uncertainty scaling is shifting the normalised distribution closer to the theoretical Gaussian distribution. Table~\ref{table:fluxcor_param} gives the flux uncertainty scaling parameters for different passbands.}
    \label{fig:fluxcordist}
\end{figure}

Figure~\ref{fig:fluxcorbias} shows the distribution of the model flux over the observed flux (see Eq.~\eqref{eq:falphai}) for all nine GAMA filters. It is visible that there is a bias between observed and model fluxes. After adding the flux corrections (see Table~\ref{table:fluxcor_param}), the distributions are centred around unity as expected for bias-free flux estimates. We emphasise that the flux corrections do not indicate that the observed fluxes themselves are biased. The bias is measured between the observed and model fluxes computed from the optimised set of templates (see Sect.~\ref{sec:templates}). The bias might also indicate that the used spectral templates are systematically biased or do not cover the entire variety of galaxy templates. The flux corrections in Table~\ref{table:fluxcor_param} are derived for the photo-z estimation in the TOPz code. For most filters, the flux bias correction is less than a few per cent, which is less than the intrinsic scatter of the difference between the model and observed fluxes.

Figure~\ref{fig:fluxcordist} shows the distribution between the model and observed fluxes normalised by the flux uncertainty. The distribution is shown before and after flux uncertainty scaling (see Eq.~\eqref{eq:fluxscaling}). The flux uncertainty scaling parameters are given in Table~\ref{table:fluxcor_param}. The theoretical expectation is that the normalised differences follow a Gaussian distribution with unit variance. Comparing the distributions before and after flux uncertainty scaling, we see that the scaling moves the distributions closer to the theoretical expectation. Altogether, the flux and flux uncertainty corrections presented in this section are derived following theoretical expectations. For the GAMA sample, the corrections given in Table~\ref{table:fluxcor_param} improve the photo-z estimates as we will demonstrate in Sect.~\ref{sec:photoz_statistics2}.

\section{Template set generation for photometric redshift estimation}
\label{sec:templates}

To estimate the photo-z of galaxies as described in Sect.~\ref{sec:topz}, we take a similar approach as \citet{2022A&A...668A...8L}. To model the redshifts of all galaxies, we need their description in the form of a spectral template set $\mathbf{T}$ that covers adequately all possible galaxy types. To generate the spectral templates of galaxies, we use the CIGALE\footnote{\url{https://cigale.lam.fr}} (Code Investigating GALaxy Emission) software version 2022.1 \citep{2019A&A...622A.103B, 2022ApJ...927..192Y}, which is a versatile tool to generate a large variety of galaxy spectra. To construct galaxy spectra, we use CIGALE by including star formation history with {\it sfh2exp} module (modelled as a combination of two exponential profiles), initial mass function, stellar evolution tracks, dust attenuation with {\it dustatt\_modified\_CF00} module, nebular emission lines, and redshifting.
\citet{2021MNRAS.502.5762C} show that emission lines with varying strength in the template spectra are improving the photo-z estimation in template-fitting methods \citep[see also][]{2006A&A...457..841I, 2013ApJ...775...93D,2014ApJ...796...60H}. Hence, we allowed CIGALE to fit a variety of emission lines in the spectra.
We did not use dust emission and AGN contribution since we use dominantly optical wavelengths (the smallest dust particles can provide thermal contribution up to $\lambda\sim3\,{\rm \mu m}$ \citep{Ryden:2021} in rare occasions, but mostly confined to $\lambda\sim120\,{\rm \mu m}$).

To generate templates with CIGALE, we used the following parameters:
\begin{itemize}
    \item Chabrier initial mass function, stellar evolution tracks:
    \begin{itemize}
        \item Metallicity $Z = 0.0001, 0.0004, 0.004, 0.008, 0.02, 0.05$
        \item separation\_age = 10 Myr
    \end{itemize}
    \item Double exponential star formation history:
    \begin{itemize}
        \item tau\_main = 0.5, 1, 2, 3, 4, 5, 6.5, 8 Gyr
        \item tau\_burst = -200, -100, 50.0, 100, 200, 600 Myr
        \item f\_burst = 0.002, 0.01, 0.03, 0.05, 0.1, 0.25, 0.5
        \item age = 8000, 5000, 3000, 2000 Myr
        \item burst\_age = 10.0, 30.0, 100.0, 500.0, 1000.0 Myr
    \end{itemize}
    \item Dust attenuation:
    \begin{itemize}
        \item Av\_ISM = 0.0, 0.1, 0.25, 0.4, 0.7, 1.0, 2.0
    \end{itemize}
    \item Nebular emission lines:
    \begin{itemize}
        \item logU = -4.0, -3.5, -3.0, -2.0, -1.5, -1.0
        \item zgas = 0.0004, 0.001, 0.004, 0.011, 0.022, 0.041
        \item ne = 100
        \item f\_esc = 0.0
        \item f\_dust = 0.0
        \item lines\_width = 300.0 $\mathrm{km}\,\mathrm{s}^{-1}$.
    \end{itemize}
\end{itemize}
We refer to the CIGALE papers and manual for a detailed description of these parameters. The output spectral templates from the CIGALE software cover the wide wavelength range from UV to IR ($10^1$--$10^9$~nm), which is sufficiently wide to cover all GAMA nine-band photometry over a large redshift range. In the current analysis, we use CIGALE to generate templates for relatively low-redshift ($z<1$) galaxies. For high-redshift galaxies, a different set of physically motivated templates is necessary \citep{2024arXiv240920519L}.

\begin{figure}
\centering
    \includegraphics{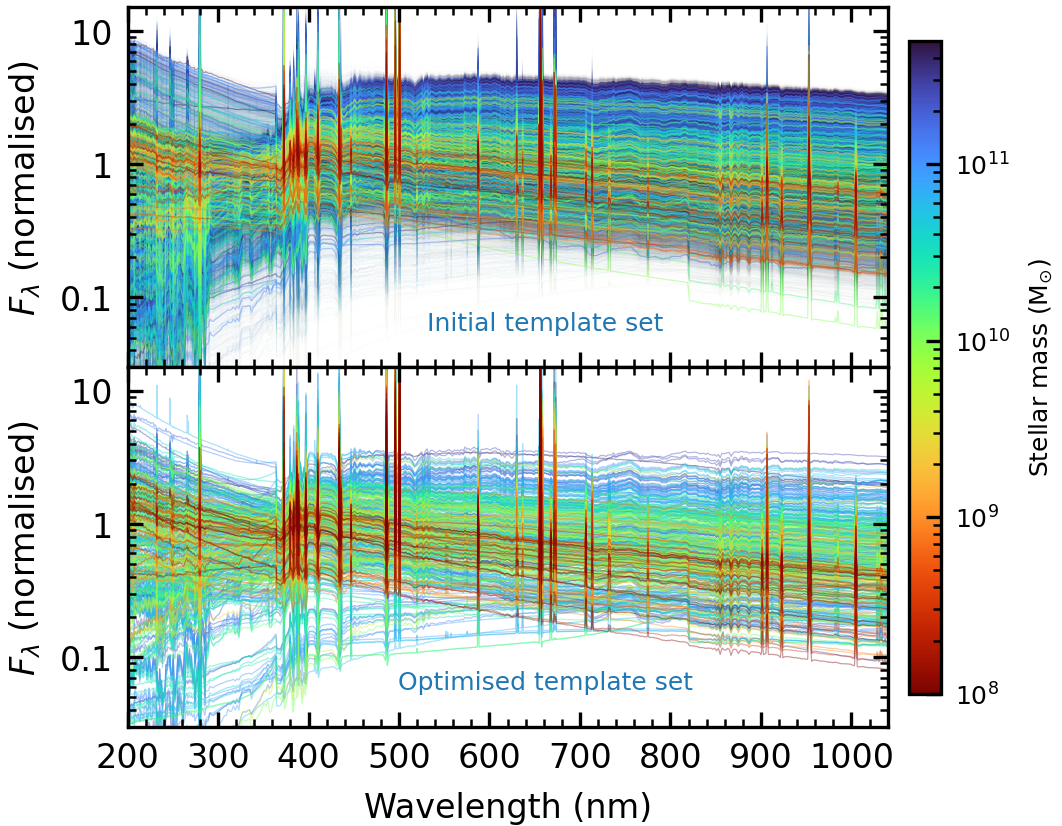}
    \caption{Spectral templates generated using the CIGALE software (see Sect.~\ref{sec:templates}). The upper panel shows all $\sim$40000 spectral templates, while the lower panel shows the selected 555 templates after template set optimisation. Template fluxes are normalised to unit stellar mass. The colour coding shows the stellar masses from the CIGALE software.}
    \label{fig:templates}
\end{figure}

Galaxy template generation for TOPz is a two-step process. In the first step, we use a spectroscopic redshift sample and CIGALE to generate a large set of templates. We fix the redshift to the spectroscopic redshift and generate a best-fitting template for each input galaxy. To keep the number of generated templates reasonable, we downsampled the entire spectroscopic GAMA catalogue to roughly 40\,000 galaxies. To avoid removing galaxies that are underrepresented in the spectroscopic catalogue, we downsampled to have uniform coverage in $\zeta$ space. Fig.~\ref{fig:gama_specz} shows the distribution of sources in the full and downsampled GAMA spectroscopic catalogue. The downsampled catalogue covers fairly uniformly the redshift and magnitude ranges in the spectroscopic sample, except for the regions where we are limited by observational data. The upper panel in Fig.~\ref{fig:templates} shows the spectral templates generated using the CIGALE software. The generated spectral templates cover a wide variety of galaxy types. In TOPz code, we assume that a single template can represent a galaxy. In general, galaxy photometry can be blended, which can be included in template-fitting methods \citep{2019MNRAS.483.2487J} but is computationally too demanding for large surveys.

\begin{figure}
\centering
    \includegraphics{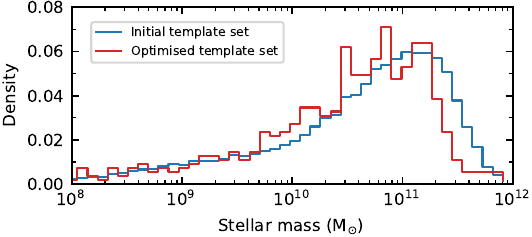}
    \caption{Normalised histogram of stellar masses for the initial set of templates (blue) and an optimised set of templates (red). Templates are generated using the CIGALE software (see Sect.~\ref{sec:templates}), and stellar masses are total stellar masses provided by the CIGALE.}
    \label{fig:stellarmass_hist}
\end{figure}

For photo-z estimation, we need a large enough template set to cover a wide range of galaxy templates. Still, at the same time, it should be sufficiently small to be computationally feasible. \citet{2023A&C....4400735F} highlight that template set optimisation improves the accuracy of the photo-z codes. To reduce the template set, we use the template set optimisation that selects only templates that maximise the number of correct photo-z. The template set optimisation is performed using the spectroscopic galaxy sample. The metric that we are optimising is the following:
\begin{equation}
    S(\mathbf{T^*}) = \frac{\sum\limits_{G \in \mathbf{G}_\mathrm{spec}} S^{\mathbf{T^*}}_G \, w_G} {\sum\limits_{G \in \mathbf{G}_\mathrm{spec}} w_G} ,
\end{equation}
where $w_G$ is a spectroscopic galaxy weight for optimisation and $S_G$ is an optimisation metric per galaxy $G$, which is defined as
\begin{equation}
    S^{\mathbf{T^*}}_G =  p^\mathrm{best}_\mathrm{peak}\left(G,{\mathbf{T^*}}\right) \mathbbm{1}\!\left\{ \zeta^{\mathrm{best}}_{\min} \!<\! \zeta_\mathrm{spec} \!<\! \zeta^{\mathrm{best}}_{\max} \right\} + 
    w_\mathrm{spec}\, p^G_{\mathbf{T^*}}(\zeta_\mathrm{spec}),
    \label{eq:SG}
\end{equation}
where $\mathbbm{1}\!\left\{\cdot\right\}$ is an indicator function that is one if the spectroscopic redshift $\zeta_\mathrm{spec}$ value is inside the best peak in the photo-z posterior and $w_\mathrm{spec}$ is a user-defined weight parameter. The optimisation metric $S_G$ maximises two aspects: the probability that the spectroscopic redshift is located inside the best peak in the photo-z posterior, and the photo-z posterior value at $\zeta_\mathrm{spec}$. The parameter $w_\mathrm{spec}$ sets the weight between these two criteria.

There is no unique way to define the weights $w_G$ for galaxies. The simplest option is to set $w_G=1.0$ for each galaxy. In TOPz code, we have also implemented an option to set the weights so that the weighted redshift distribution is uniform in $\zeta$ space. This will suppress the inhomogeneities in the spectroscopic redshift sample so that the optimised template set $\mathbf{T}_\mathrm{opt}$ is not only optimised for overrepresented redshift ranges.

In Eq.~\eqref{eq:SG}, the $\mathbf{T}^*$ is an arbitrary set of templates drawn from the full template set $\mathbf{T}$ that was generated using the CIGALE software. The optimised set of templates $\mathbf{T}_\mathrm{opt}$ is defined as
\begin{equation}
    \mathbf{T}_\mathrm{opt} = \argmax_{{\mathbf{T}^*} \subseteq \mathbf{T}} \left\{ S(\mathbf{T^*}) \right\}.
\end{equation}
The optimisation is carried out using Metropolis-Hastings algorithms with birth and death moves. The Metropolis-Hastings algorithm will simultaneously optimise the best number of templates and the actual set of templates.

For maximal photo-z performance, the template set optimisation can be carried out iteratively. The first round of optimisation defines a set of templates that finds a good photo-z for the majority of the galaxies. Using the optimised template set $\mathbf{T}_\mathrm{opt}$, we can identify the set of galaxies where photo-z deviates significantly from the spectroscopic redshift, i.e. the catastrophic outlier sample. This set of catastrophic outlier galaxies can be used to generate additional templates with the CIGALE software. Combining the original optimised set $\mathbf{T}_\mathrm{opt}$ with the newly generated templates for catastrophic outliers, we have a new full set of templates $\mathbf{T}$ that can be used to find the new optimised set $\mathbf{T}_\mathrm{opt}$. There is no guarantee that this process will converge. In practical applications, every iteration of the template set optimisation reduces the number of catastrophic outliers and increases the general performance of the photo-z estimation. In the current paper, for the GAMA sample, we used three iterations. In every iteration, we selected approximately 10\,000 galaxies with bad photo-z estimates and generated new templates with CIGALE that were added to the template set $\mathbf{T}$.

\begin{figure}
\centering
    \includegraphics[width=\columnwidth]{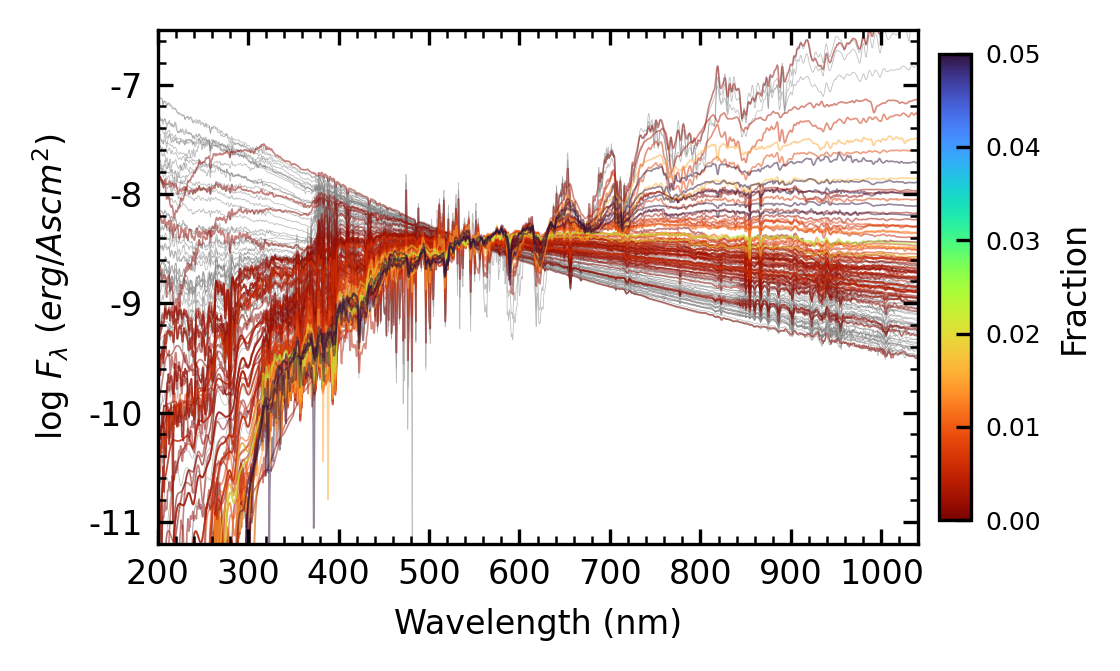}
    \caption{All 131 stellar templates taken from the TRDS Pickles Atlas. The colour coding shows the fraction of objects that have been assigned a $p_\mathrm{star}>0.5$ for that specific stellar template, despite all of the objects being designated as galaxies in the GAMA catalogue. Dark blue lines show the stellar templates that are most often assigned to galaxies, while templates shown in grey are stellar templates that are not confused with galaxies in the GAMA data set.}
    \label{fig:pickles}
\end{figure}

The lower panel of Fig.~\ref{fig:templates} shows the final set of 555 spectral templates used for the photo-z estimation. Compared with the initial template set (upper panel), the optimised set covers all possible galaxy types uniformly. Fig.~\ref{fig:stellarmass_hist} shows the distribution of stellar masses for the initial and optimised set of templates. The distributions cover the entire stellar mass range, confirming that the optimised template set covers all possible galaxy types. There are slightly fewer templates in the optimised set for the most massive templates (see Figs.~\ref{fig:templates} and \ref{fig:stellarmass_hist}). This is expected since most massive galaxies are ellipticals whose spectra are more similar to each other than the spectra of spiral galaxies are to one another. This similarity arises from their homogeneous stellar populations, lack of star formation, and minimal dust and gas content. Hence, fewer such templates are sufficient for photo-z estimation in the TOPz code.

\section{Star-galaxy identification with template fitting}
\label{sec:stargal}

Different types of astronomical objects have different spectral energy distributions (SED). The SED of galaxies is generally different from the SED of stellar objects. Comparing the model spectral templates with the SED of observed objects, it is possible to find the template that best matches the observed fluxes of an object. Hence, including the stellar templates in the set of model templates, it is possible to perform a simple star-galaxy identification using template fitting.

The stellar templates included in the TOPz code are taken from the TRDS Pickles Atlas\footnote{\url{https://www.stsci.edu/hst/instrumentation/reference-data-for-calibration-and-tools/astronomical-catalogs/pickles-atlas}} \citep{1998PASP..110..863P}. It provides comprehensive spectral coverage of 131 flux-calibrated stellar spectra, which encompass all typical spectral types and luminosity classes at solar abundance, as well as metal-weak and metal-rich F-K dwarf and G-K giant stars. The variety of stellar templates in the Pickles library is shown in Fig.~\ref{fig:pickles}.

We treat the Pickles stellar templates differently from the galaxy templates for the template fitting. Stellar templates are never redshifted and are all located at redshift zero. The total number of stellar templates is significantly smaller than that of redshifted galaxy templates. Hence, including stellar templates has only a small impact on the total computation time.

To perform the star-galaxy identification, we estimate the probability that a given object is a star given the observed fluxes $F$. The probability $p_\mathrm{star}$ is defined as
\begin{equation}
    p_\mathrm{star} = \frac{p^{\max}_\mathrm{star}}{p^{\max}_\mathrm{star} + p^{\max}_\mathrm{gal}} ,
\end{equation}
where
\begin{eqnarray}
    p^{\max}_\mathrm{star} &=& \max_{T \in \mathbf{T}_\mathrm{star}} p(F\,|\,T) , \\
    p^{\max}_\mathrm{gal} &=& \max_{\zeta,\,T \in \mathbf{T}_\mathrm{gal}} p(F\,|\,\zeta,T) .
\end{eqnarray}
The probability $p_\mathrm{star}$ does not depend directly on the number of used templates. However, the $p_\mathrm{star}$ depends on the coverage of different stellar types in the input library. Below, we will also use an extended stellar library to analyse this aspect specifically. In computing the $p_\mathrm{star}$, we compare the best-fitting galaxy template against the best-fitting stellar template. Additionally, we ignore the prior while computing the $p_\mathrm{star}$. Consequently, the $p_\mathrm{star}$ only depends on the observed object SED and gives the probabilistic estimate that the object has a stellar SED. In the last equations, $\mathbf{T}_\mathrm{star}$ and $\mathbf{T}_\mathrm{gal}$ refer to stellar and galaxy template libraries, respectively.

\begin{figure}
    \centering
    \includegraphics[width=\columnwidth]{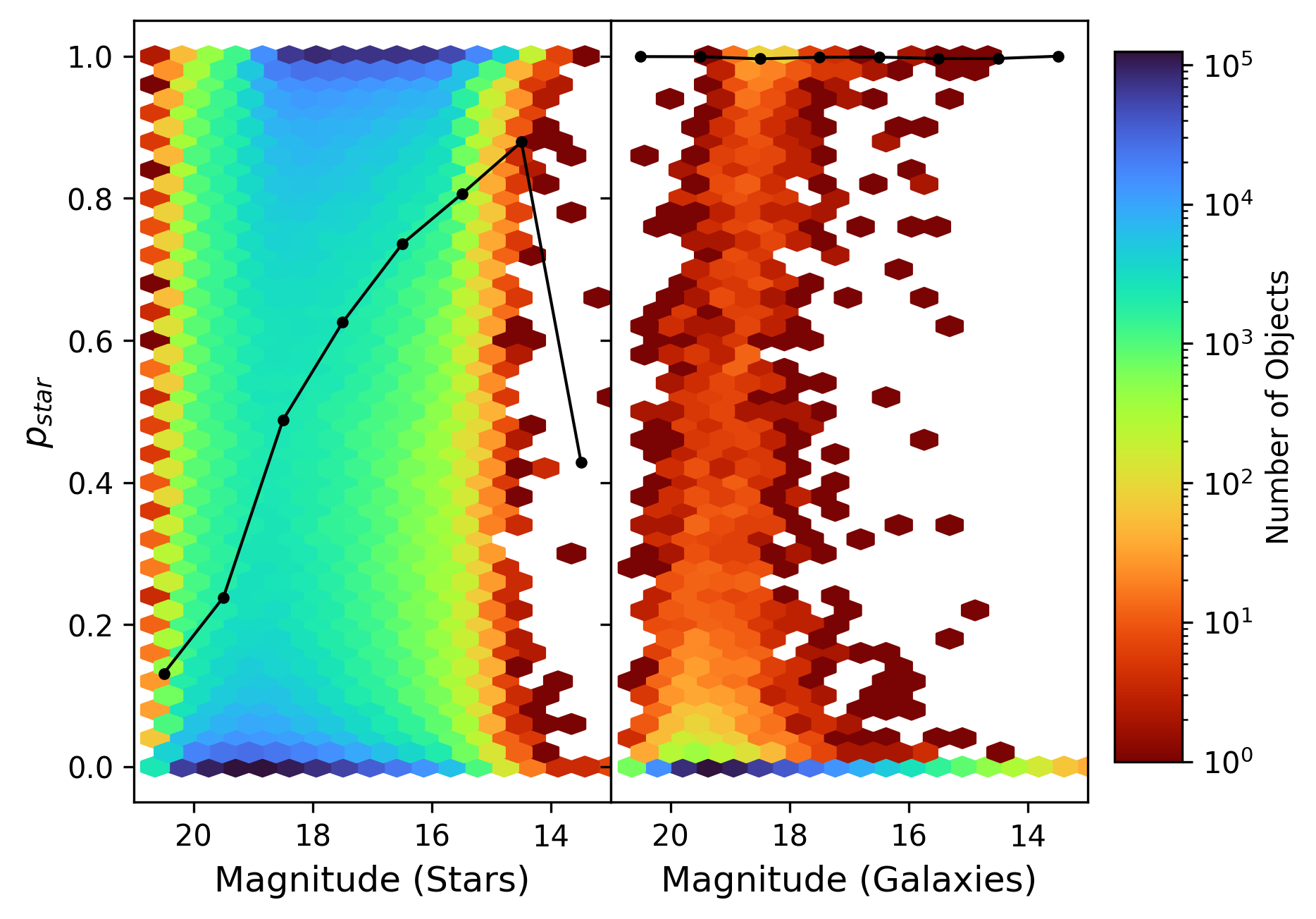}
    \caption{Objects from the GAMA catalogue plotted on the magnitude-$p_\mathrm{star}$ plane and separated depending on whether they are designated as stars (left panel) or galaxies (right panel) in the GAMA catalogue. There are 620\,131 objects in the left panel and 481\,130 objects in the right panel. The colour coding shows the number of objects in each section of the panels. The black lines in each panel show the fraction of objects that are correctly identified by TOPz ($p_\mathrm{star} \geq 0.5$ for stars, and $p_\mathrm{star} < 0.5$ for galaxies) per magnitude bin. For galaxies, the fraction of correctly classified objects over all magnitude bins is 99.8 per cent.}
    \label{fig:Pstar_vs_mag}
\end{figure}

\begin{figure}
\centering
    \includegraphics[width=\columnwidth]{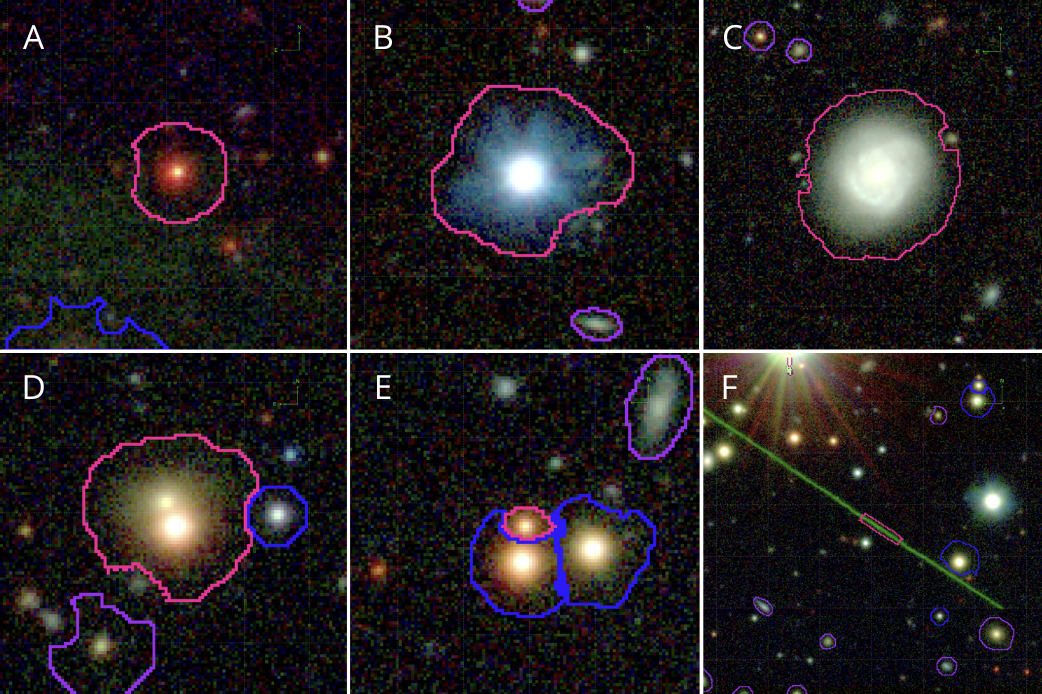}
    \caption{Example of objects classified as galaxies in the GAMA database but classified as stars by TOPz, with $p_{\text{star}} \approx 1$. Pink segments mark the misclassified objects. Purple segments indicate galaxies and blue segments indicate stars (according to the GAMA classification). In certain cases, the GAMA catalogue assigns galaxy classifications to sources that are evidently stars or image artefacts:
\textbf{A}) a compact red galaxy classified as an M-type star in TOPz; 
\textbf{B}) a bright blue star;
\textbf{C}) a luminous spiral galaxy; 
\textbf{D}) two blended sources treated as a single object; 
\textbf{E}) a small, faint object near several larger, brighter sources; 
\textbf{F}) a satellite trail.}
    \label{fig:stargalexample}
\end{figure}

\begin{figure}
\centering
    \includegraphics[width=\columnwidth]{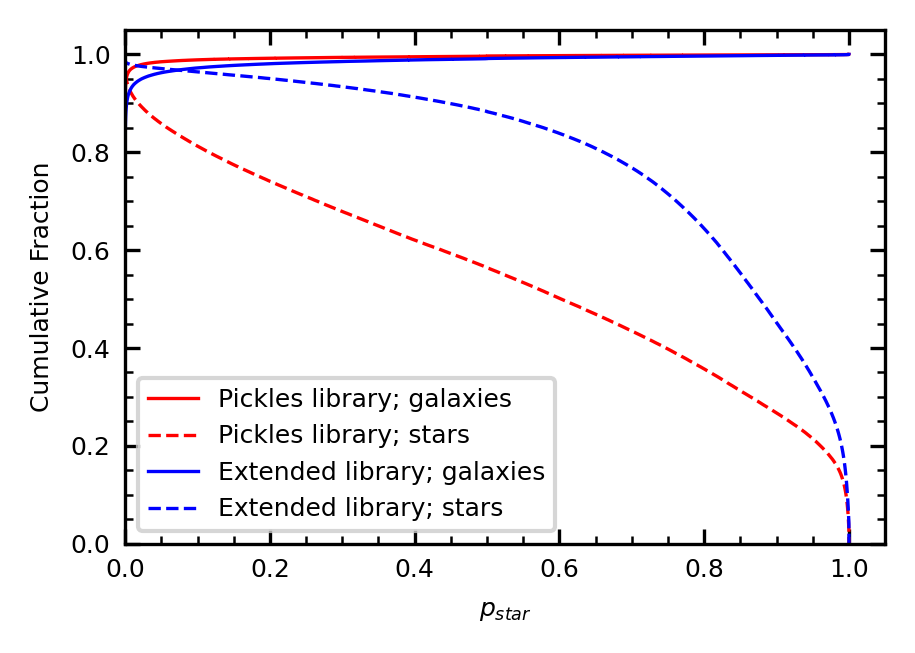}
    \caption{Cumulative fractions of object classifications using the Pickles stellar library (red curves) and the extended stellar library (blue curves). The solid lines show the forward cumulative distributions of galaxies and the dashed lines show the reversed cumulative distributions of stars. Classification of stars and galaxies is taken from the GAMA database.}
    \label{fig:stargal}
\end{figure}

In Fig.~\ref{fig:pickles} we show in colour the Pickles stellar templates that are most often confused with galaxies. The colour coding shows the fraction of how many times each stellar template best matches an object compared with all stellar templates. We see that redder stellar templates are most often confused with galaxies. In Fig.~\ref{fig:Pstar_vs_mag} we show the $p_\mathrm{star}$ probability as a function of object magnitude for stars and galaxies as classified by the GAMA survey (see Sect.~\ref{sec:gama}). For stars in the GAMA database, many objects are classified as galaxies in the TOPz. However, there are only a few galaxies where the $p_\mathrm{star}$ probability is larger than 0.5. Hence, very few galaxies (less than 0.2 per cent) are misclassified by TOPz. When visually investigating some of these misclassified galaxies, we find that those with $p_\mathrm{star}$ probabilities close to 1 are very small, very red objects that are most often assigned as M-type stars by TOPz. These galaxies are roughly split equally between being visually isolated and surrounded by other objects in the sky. Some of those objects are genuine Milky Way stars projected next to a background galaxy. In Fig.~\ref{fig:stargalexample}, we provide examples of GAMA galaxies classified as stars in the TOPz code.

\begin{figure}
\centering
    \includegraphics[width=\columnwidth]{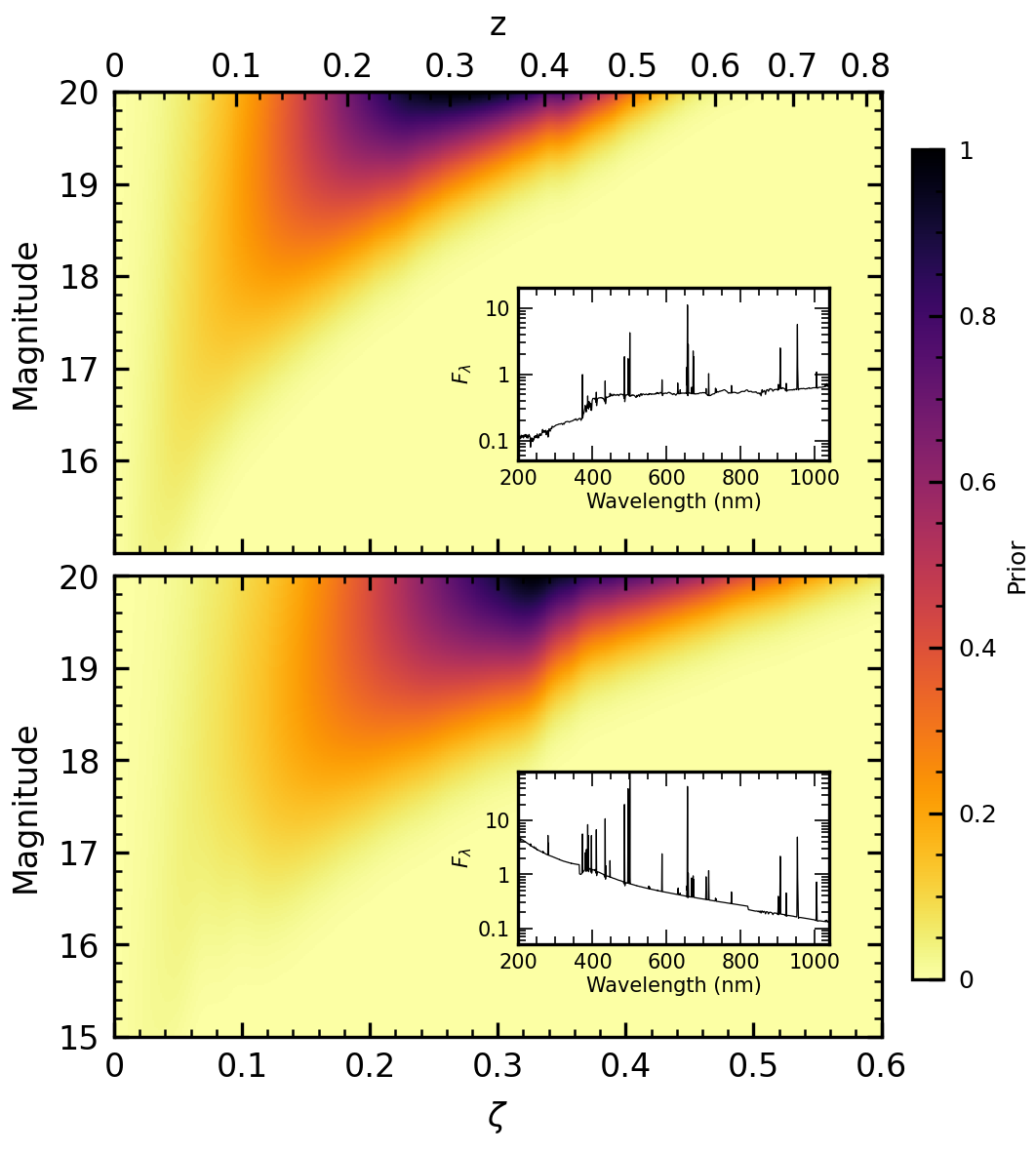}
    \caption{Redshift prior dependence on redshift $\zeta$ and apparent galaxy magnitude for two different templates (upper and lower panel). The inset panels show the template spectra. The two templates were selected arbitrarily from the optimised template set (see Fig.~\ref{fig:templates}) and serve as examples. The priors are arbitrarily normalised. The decrease of prior towards redshift zero is because of the volume prior, while the decrease at larger redshifts is due to the luminosity function prior. The differences in spectral templates affect the higher redshift end due to different k-corrections. The slight deviation from a smooth prior is caused by the features (e.g. emission lines) in the spectral templates.}
    \label{fig:priormag}
\end{figure}

\begin{figure}
\centering
    \includegraphics[width=\columnwidth]{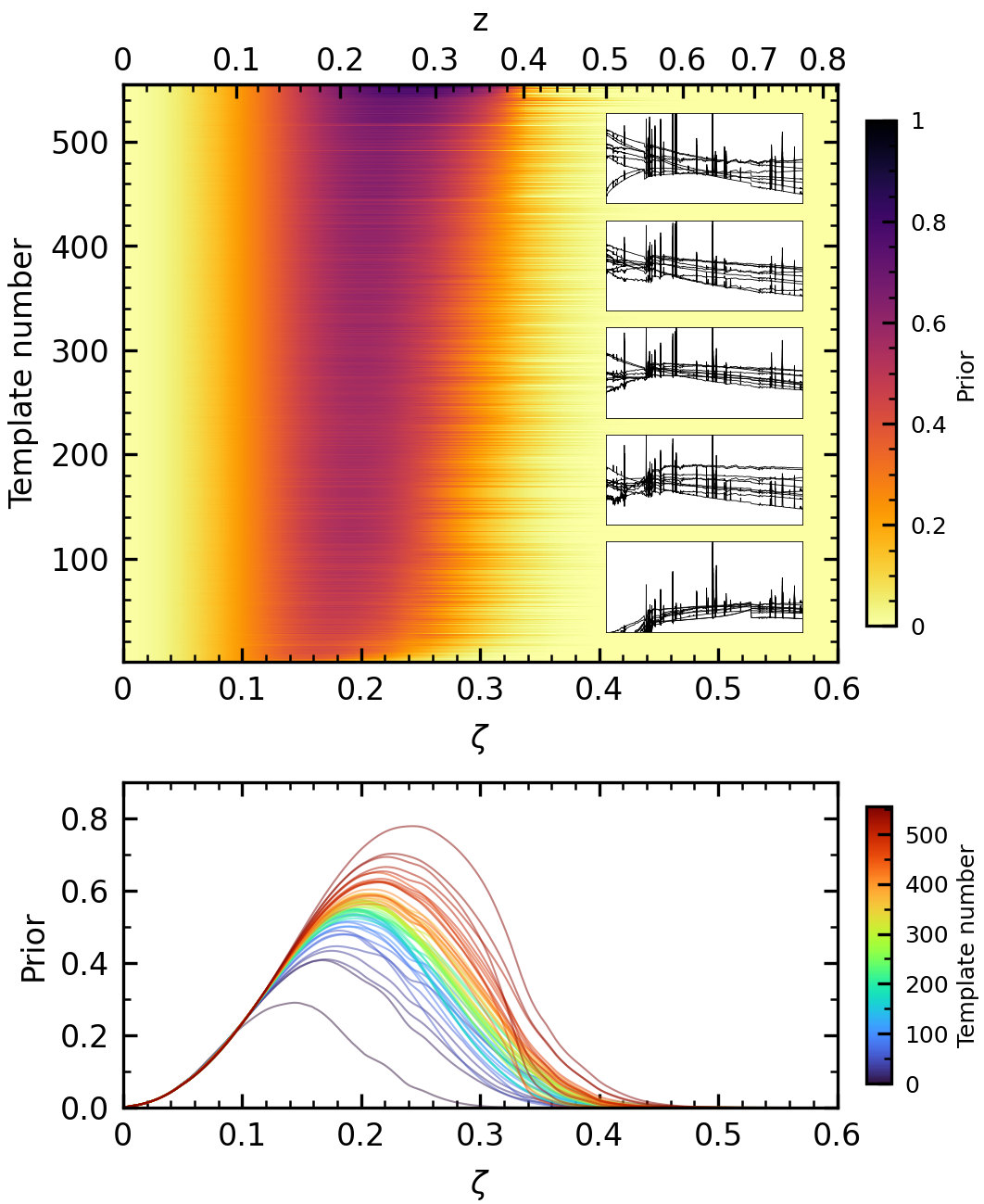}
    \caption{\textbf{Upper panel:} Redshift prior for a fixed apparent magnitude ($m_Z=19\,$mag) for all templates in the optimised template set (see lower panel in Fig.~\ref{fig:templates}). The templates are ordered based on the prior weighted mean redshift. The inset panels show how the template spectra change from top to bottom. \textbf{Lower panel:} Redshift prior for different templates drawn from the upper panel.}
    \label{fig:priormatrix}
\end{figure}

To test the star-galaxy classification dependence on the choice of the used stellar library, we extended the original Pickles library with additional stellar templates. We added Kurucz ODFNEW/NOVER theoretical stellar spectra \citep{1997A&A...318..841C, 2003IAUS..210P.A20C} and added white dwarf models from \citet{2010MmSAI..81..921K}. In total, we have 9287 spectra in our extended stellar library. Since the GAMA field is located in the Milky Way halo, we will ignore the interstellar reddening, which is mostly a concern in the Milky Way disc region. Figure~\ref{fig:stargal} shows the cumulative classification probability for objects classified as stars or galaxies in the GAMA database. For galaxies, the extended stellar library makes a small difference, while for stars, the extended stellar library significantly reduces the fraction of wrongly classified objects.

We emphasise that the test was carried out using relatively bright sources in the GAMA photometric catalogue. The confusion between stellar and galaxy templates is significantly increased for fainter targets with larger uncertainties for observed fluxes. Hence, for fainter sources, the simple SED classification is only a rough indication of the type and cannot be used as a reliable star-galaxy classification. For accurate star-galaxy classification, the object morphology (e.g. an indication of a point-like source) should be included as well \citep[e.g.][]{2024MNRAS.535.2129C}.

\section{Luminosity function prior for photometric redshifts}
\label{sec:prior}

\begin{figure*}
\centering
    \includegraphics[]{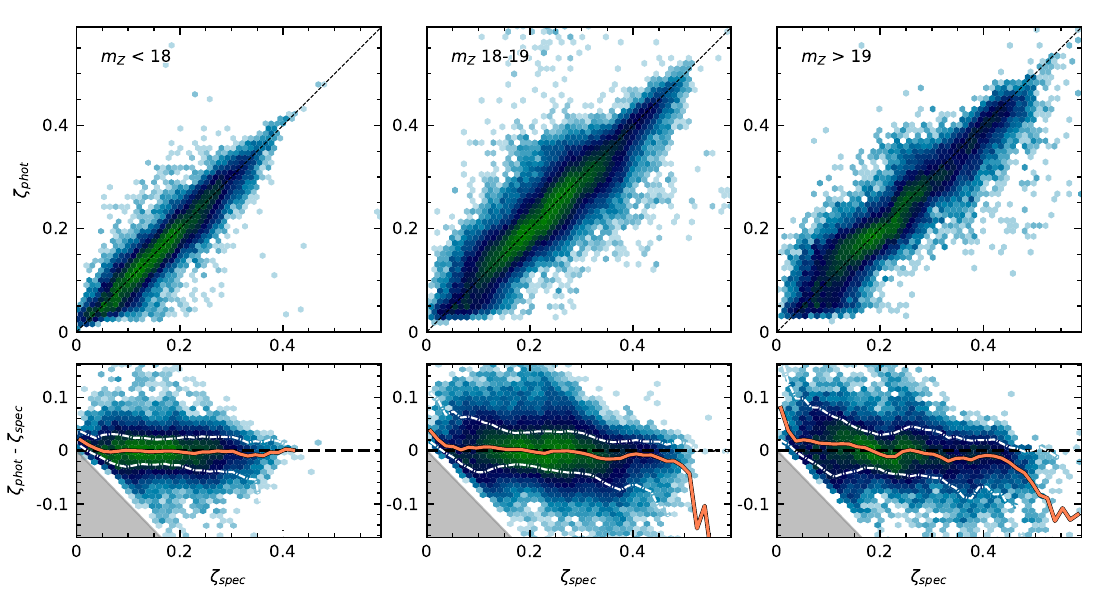}
    \caption{Correspondence between spectroscopic redshifts and TOPz photo-z estimations. \textbf{Upper panel}: The colour of the hexagons denotes the number of galaxies on a logarithmic scale. The one-to-one relation is drawn with a diagonal dashed black line. See Fig.~\ref{fig:dz_distribution} for the distribution around the one-to-one line. From left to right, the three panels include galaxies in different $Z$-band magnitude ranges: $m_Z < 18$, $18 < m_Z < 19$ and $m_Z > 19$, respectively. \textbf{Lower panel}: Photo-z accuracy as a function of spectroscopic redshift. The three panels are comprised of galaxies at the same magnitude ranges as on the top panels, and hexagon colours are similar. The coral line shows the median accuracy in different spectroscopic redshift bins, and the lower and upper white dash-dot lines indicate the 10\% and 90\% data quantiles, respectively. The grey area at the bottom-left of each panel indicates an exclusion zone where $\zeta_\mathrm{phot}$ would be negative, causing the increase in the median (coral line) when approaching redshift zero.}
    \label{fig:zspec_vs_zphot}
\end{figure*}

Photo-z prior $p(\zeta,T\,|\,m_0)$ gives a likelihood that a galaxy with a total magnitude $m_0$ is located at redshift $\zeta$ and has a spectral type $T$. The prior $p(\zeta,T\,|\,m_0)$ can be further expressed as
\begin{equation}
    p(\zeta,T\,|\,m_0) = p(T\,|\,m_0) p(\zeta\,|\,m_0,T),
\end{equation}
where $p(\zeta\,|\,m_0,T)$ is the redshift distribution of galaxies with total magnitude $m_0$ and spectral template $T$, and $p(T\,|\,m_0)$ defines the prior for different spectral templates that in general depends on magnitude $m_0$. The $p(T\,|\,m_0)$ depends on the template set $\mathbf{T}$ and should be independently estimated for a given template set. For simplicity, we use a non-informative prior in a given paper and set $p(T\,|\,m_0) = \mathrm{const}$. This is justified since our optimised set of templates is selected based on the observed data (see Sect.~\ref{sec:templates}).

For template-fitting photo-z codes, \citet{2015ApJ...801...20T} emphasises the importance of physically motivated priors. Hence, in the current paper, the redshift distribution prior $p(\zeta\,|\,m_0,T)$ is defined using a luminosity function of galaxies in a fixed passband. The prior is defined as
\begin{equation}
    p(\zeta\,|\,m_0,T) \propto \Phi(L\,|\,\zeta) V(\zeta),
\end{equation}
where $\Phi(L\,|\,\zeta)$ is the galaxy luminosity function at redshift $\zeta$, where $L$ is the absolute luminosity of a galaxy. We assume that the number count of galaxies (luminosity function normalisation) does not depend on the redshift. The term $V(\zeta)$ defines the redshift-dependent volume term in the prior.

We employ the Schechter luminosity function that allows us to write
\begin{equation}
    \Phi(L\,|\,\zeta) = \Phi(L\,|\,\zeta,L^\star,\beta) \propto
    \left(\frac{L}{L^\star}\right)^\beta \exp{ \left(\frac{L}{L^\star}\right) },
\end{equation}
where the luminosity function depends on characteristic luminosity $L^\star$ and power law index $\beta$. To account for the evolution correction of galaxies, the characteristic luminosity $L^\star$ is defined as
\begin{equation}
    L^\star \equiv L^\star(\zeta) = L_0^\star \cdot 10^{ -0.4\left( z Q \right) },
\end{equation}
where $Q<0$ means that galaxies were brighter in the past \citep[see e.g.][]{2003ApJ...592..819B, 2014A&A...566A...1T}. While specifying the luminosity function parameters $L_0^\star$, $\beta$, and $Q$, we can analytically calculate the redshift-dependent prior for a galaxy with absolute luminosity $L$. The luminosity function is defined for a fixed passband $\alpha_\mathrm{LF}$ and the corresponding absolute luminosity $L$ should be measured in the same passband at the rest frame. To convert the observed total magnitude $m_0$ measured in passband $\alpha_\mathrm{tot}$ into absolute luminosity $L$ measured in passband $\alpha_\mathrm{LF}$ we do the following
\begin{equation}
    L_{\alpha_\mathrm{LF}}(\zeta,T) = f_{\nu,0}10^{-0.4 m_0} \frac{D_\mathrm{L}(\zeta)}{10\,\mathrm{pc}} \frac{F_{T\alpha_\mathrm{LF}}(\zeta_\mathrm{LF})}{F_{T\alpha_\mathrm{tot}}(\zeta)},
\end{equation}
where $f_{\nu,0}$ is zero point of the magnitude system. If $m_0$ is given in AB magnitude system then $f_{\nu,0}=3630.78\,\mathrm{Jy}$. The $D_\mathrm{L}(\zeta)$ is a luminosity distance at redshift $\zeta$ computed assuming a specified cosmology. The last term accounts for the k-correction and the difference between the passbands of total magnitude and luminosity function. It is estimated for each template $T$ individually using model fluxes computed with Eq.~\eqref{eq:model_flux}.

Figure~\ref{fig:priormag} shows the used prior for two different templates. The prior dependence on observed magnitude is defined by the luminosity function prior. The prior dependence on redshift is a combination of a volume prior and a luminosity function prior. The decrease of prior towards redshift zero comes from the volume prior, while the decrease at larger redshift comes from the luminosity function prior. The general shape of the chosen prior is similar to the previously proposed general analytical prior \citep{2000ApJ...536..571B}.

Figure~\ref{fig:priormatrix} shows the redshift prior for the complete set of optimised templates at a fixed apparent magnitude. At lower redshifts, the prior is roughly the same for all the templates. At higher redshifts, the prior for different templates varies slightly due to the distinct k-corrections associated with each template. In general, the redshift range, where the prior is significant, is roughly the same for all templates. Hence, the shape of the luminosity function is more important than the spectral differences between the templates.

\section{TOPz performance using GAMA sample}
\label{sec:topz_gama}

To test the performance and accuracy of the TOPz code, we have applied the TOPz algorithm described in the previous sections to the GAMA spectroscopic sample (see Sect.~\ref{sec:gama}). Table~\ref{table:topz_param} lists all the parameters defined in previous sections and used for the GAMA data. Applied photometric flux corrections are given in Table~\ref{table:fluxcor_param} and described in Sect.~\ref{sec:fluxcorr}. Sect.~\ref{sec:prior} presents the prior used in the current analysis. In this section, we will present how well the TOPz algorithm is able to estimate the photo-z and redshift uncertainties of galaxies compared to the spectroscopic redshift sample.

\begin{figure}
\centering
    \includegraphics[]{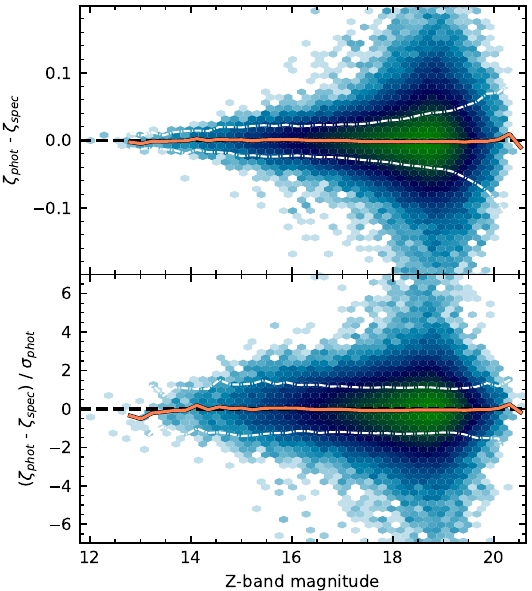}
    \caption{Photo-z accuracy dependence on total observed magnitude $m_0$. The hexagon colours indicate the number of galaxies on a logarithmic scale. The coral line shows the median difference, and the white dash-dot lines indicate the 10\% and 90\% quantiles. \textbf{Upper panel}: Photo-z accuracy for single galaxies. \textbf{Lower panel}: Photo-z accuracy normalised by the TOPz photo-z uncertainty ${\sigma}_\mathrm{phot}$ value (see Eq.~\eqref{eq:zeta_sigma}).}
    \label{fig:dz_vs_mag}
\end{figure}

\begin{figure}
\centering
    \includegraphics[]{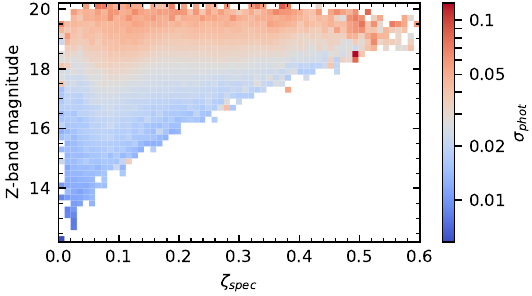}
    \caption{Photo-z peak width depending on the galaxy's spectroscopic redshift and $Z$-band magnitude. For clarity, each magnitude and redshift bin shows the average peak width of the galaxies in that bin.}
    \label{fig:zspec_mag_sigma}
\end{figure}

\begin{table}
\caption{TOPz parameters for the GAMA data set.}
\label{table:topz_param}
\centering
\begin{tabular}{l c l}
\hline\hline
Parameter & Value & Section \\
\hline
   $f_\mathrm{peak}$ & 0.005 & Sect.~\ref{sect:topz_output} \\
   $w_\mathrm{spek}$ & 10.0 & Sect.~\ref{sec:templates} \\
   $\mathrm{delta}\_\zeta$ & 0.1 & Sect.~\ref{sec:fluxcorr} \\
   $\mathrm{delta}\_\mathrm{flux}$ & 3.0 & Sect.~\ref{sec:fluxcorr} \\
   $\mathrm{SN}\_\mathrm{limit}$ & 5.0 & Sect.~\ref{sec:fluxcorr} \\
   $f_\mathrm{mod}^\mathrm{lim}$ & 0.5 & Sect.~\ref{sec:fluxcorr} \\
   $f_\mathrm{mod}^\mathrm{diff}$ & 0.01 & Sect.~\ref{sec:fluxcorr} \\
   $L_0^\star$ & -21.4 & Sect.~\ref{sec:prior} \\
   $\beta$ & -0.7 & Sect.~\ref{sec:prior} \\
   $Q$ & -0.6 & Sect.~\ref{sec:prior} \\
\hline
\end{tabular}
\end{table}

\subsection{Photometric redshifts for the GAMA sample}
\label{sec:photoz_gama}

Figure~\ref{fig:zspec_vs_zphot} shows the comparison between the spectroscopic redshifts and photo-z for three different magnitude bins. The galaxy redshifts are recovered visually well without a clear bias in photo-z estimates. All magnitude bins behave similarly, while the scatter is larger for fainter galaxies. This is expected since photo-z accuracy depends on the accuracy of the input photometry. Table~\ref{table:topz_photoz_params} quantifies numerically the TOPz photo-z estimates. Fig.~\ref{fig:dz_vs_mag} shows the photo-z accuracy as a function of galaxy total magnitude. The median of the estimated photo-z is close to zero, indicating that the TOPz estimated photo-z is nearly bias-free. The scatter around the median strongly depends on the galaxy's total magnitude due to the increased uncertainty of photometric flux measurements. The lower panel in Fig.~\ref{fig:dz_vs_mag} shows the normalised difference between spectroscopic redshift and photo-z estimates; the redshift difference is normalised with the TOPz photo-z uncertainty. The normalised difference is independent of apparent magnitude, indicating that the photo-z uncertainties adequately take into account the observed flux uncertainties. In Fig.~\ref{fig:zspec_mag_sigma} we show how the photo-z uncertainty depends on magnitude and spectroscopic redshift. As expected, the uncertainty dominantly depends on the magnitude (flux uncertainty). In Sect.~\ref{sect:photoz_pdf_test}, we specifically analyse the accuracy of the photo-z probability distribution functions (posteriors).

\begin{figure}
\centering
    \includegraphics[]{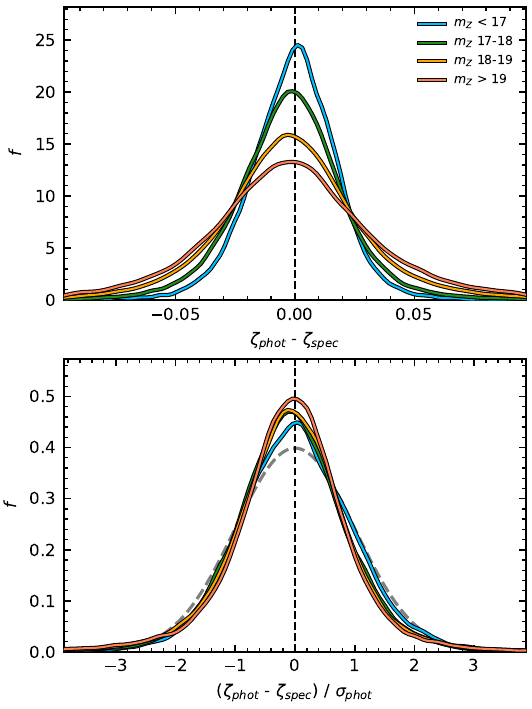}
    \caption{Photo-z accuracy in multiple magnitude ranges. \textbf{Upper panel}: The colours show the distribution of photo-z accuracy, which is given as the difference between the spectroscopic and photo-z estimates. The dashed black line is drawn at zero to identify any biases in the distributions. \textbf{Lower panel}: The distribution of redshift accuracies that are normalised by the width of their photo-z peak computed using Eq.~\eqref{eq:zeta_sigma}. The dashed grey line shows a Gaussian distribution with a standard deviation of one.}
    \label{fig:dz_distribution}
\end{figure}

\begin{figure}
\centering
    \includegraphics[]{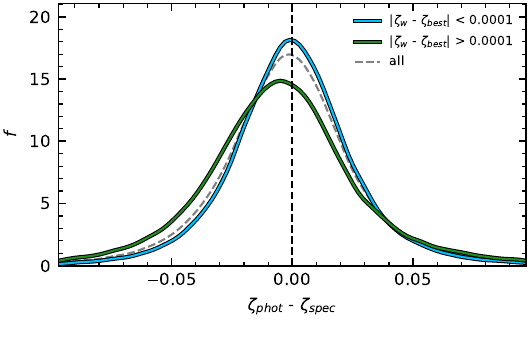}
    \caption{Photo-z accuracy distribution for two TOPz subsets. The blue line shows the subset where $\zeta_\mathrm{best}$ and $\zeta_\mathrm{w}$ are closer than $0.0001$, and the green line shows the subset with the rest of the galaxies. The dashed grey line is the distribution for the whole sample. The value $0.0001$ was chosen so the subsets would contain roughly comparable numbers of galaxies.}
    \label{fig:w_vs_best}
\end{figure}

\begin{figure}
\centering
    \includegraphics[]{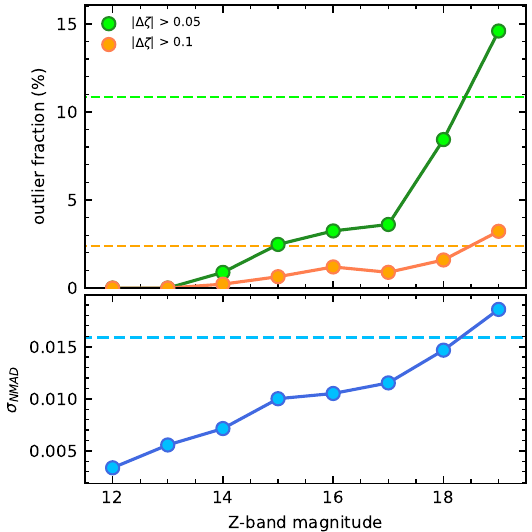}
    \caption{TOPz photo-z statistics as a function of the apparent $Z$-band magnitude. Photo-z accuracy is defined as $\sigma_\mathrm{NMAD}$ (see Sect.~\ref{sec:photoz_statistics} and Eq.~\eqref{eq:sigma_nmad}). Outlier fraction is defined using redshift $\zeta$ absolute deviation from spectroscopic redshift for two different $\Delta\zeta$ thresholds: 0.05 and 0.1. \textbf{Upper panel}: Outlier fraction as a function of apparent magnitude. The outlier fractions for two different thresholds are marked in green and orange colours. The coloured dashed lines show the outlier fractions over the whole galaxy sample. \textbf{Lower panel}: Photo-z accuracy distribution widths as described by the ${\sigma}_\mathrm{NMAD}$ value. The dashed line shows the ${\sigma}_\mathrm{NMAD}$ value calculated over the whole galaxy set.}
    \label{fig:outlier_sigmanmad_vs_mag}
\end{figure}

\begin{figure}
\centering
    \includegraphics[]{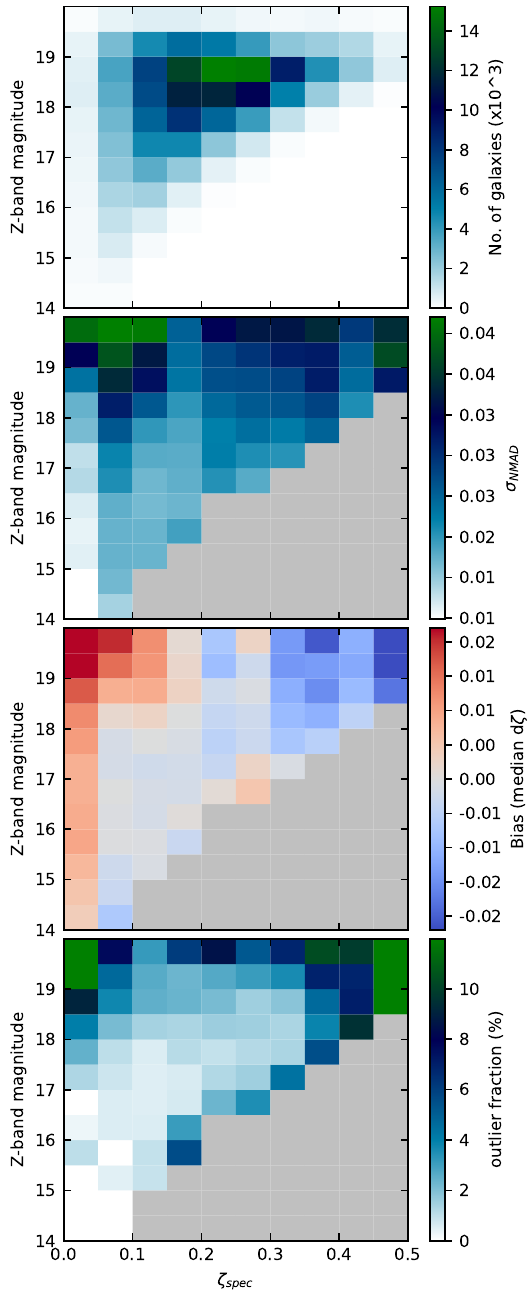}
    \caption{Photo-z estimation parameter map on a magnitude - $\zeta_\mathrm{spec}$ grid. The top panel shows the number of galaxies in each bin. The three parameters from the second to the bottom panel are photo-z accuracy distribution width ($\sigma_\mathrm{NMAD}$ value), the bias of that distribution as the median value of the photometric accuracy, and the outlier rate as the number of galaxies with photometric accuracy worse than 0.1. Bins with fewer than 100 galaxies are drawn as grey areas.}
    \label{fig:zspec_mag_grid}
\end{figure}

Figure~\ref{fig:dz_distribution} shows the distribution of differences between the photometric and spectroscopic redshifts. It is visible that the peak of the distribution is around zero for all magnitude bins. However, the width of the distribution depends on the magnitude and, as expected, is narrower for brighter objects. The lower panel in Fig.~\ref{fig:dz_distribution} shows the distribution differences between spectroscopic and photo-z normalised by the photo-z uncertainty. The dashed grey line shows the normalised Gaussian distribution with a standard deviation of one. In general, the width of photo-z differences is close to the normal distribution, indicating that the photo-z uncertainties are reasonable and that they are not over- or under-estimated. This is valid for all magnitude bins, showing that the photo-z uncertainties are statistically correct for all galaxies, independent of their observed magnitude. Additionally, we see that the distribution of observed differences peaks slightly more than the expected normal distribution. This might indicate that the photo-z errors are slightly overestimated or that photo-z uncertainties do not follow exactly the theoretical normal distribution. The latter is possible since the template fitting is based on physical galaxy templates, which might improve the photo-z accuracy. More detailed analysis of this should be based on the synthetic galaxies, which require extensive analysis and are beyond the scope of the current paper.

The TOPz code provides the photo-z estimation for up to three best peaks in the posterior distribution. In the GAMA photometric sample, 99 per cent of galaxies have only one dominant peak where $p^\mathrm{best}_\mathrm{peak} > 0.5$. The second and third peaks are usually very weak in the posterior distribution. About 5 per cent of objects have $p^\mathrm{alt1}_\mathrm{peak} > 0.05$ and only 0.5 per cent of objects have $p^\mathrm{alt2}_\mathrm{peak} > 0.05$. In some cases, the best peak dominates in the photo-z posterior, but the weaker peaks might be closer to the true spectroscopic redshift. In the GAMA spectroscopic sample, for 2 per cent of objects, the second best peak is closer to the true redshift than the best peak. The third peak is closest to the true redshift only for 0.3 per cent of objects. Hence, for general analysis, considering only the dominant peak is sufficient.

In Fig.~\ref{fig:w_vs_best}, we divided the GAMA sample into two roughly equal subsamples based on the difference between the $|\zeta_\mathrm{peak}-\zeta_\mathrm{w}|$. The figure illustrates that the photo-z estimates are statistically more accurate when the difference is small. However, the difference alone is not a good measure to quantify the accuracy of individual photo-z estimates.

\subsection{Photo-z outliers, bias, and $\sigma_\mathrm{NMAD}$}
\label{sec:photoz_statistics}

\begin{table}
\caption{TOPz photo-z parameters in three magnitude bins from Fig.~\ref{fig:zspec_vs_zphot}.}
\label{table:topz_photoz_params}
\centering
\begin{tabular}{l c c c}
\hline\hline
 & $m_Z$ < 18 & 18 < $m_Z$ < 19 & $m_Z$ > 19  \\
\hline
\noalign{\vspace{2pt}}
    BIAS  &  & &\\
   median($dz$)  & -0.001094 &  -0.001811 & -0.002050\\
   median($d\zeta$)  & -0.001095 &  -0.001813 & -0.002052\\[4pt]
   OUTLIERS  &  & & \\
   |$dz$| > 0.1  & 1.04\% &  2.44\% & 4.99\%\\
   |$dz$| > 0.15  & 0.48\% &  1.00\%  & 2.05\%\\
   |$d\zeta$| > 0.1  & 1.04\% &  2.38\% & 4.97\% \\
   |$d\zeta$| > 0.15  & 0.45\% &  0.97\% & 2.05\% \\
   |$d\zeta$|$ / \sigma_\mathrm{phot}$ > 3  & 1.84\% &  2.37\% & 3.35\%\\[4pt]
   PRECISION  &  & & \\
   $\sigma$($dz$) & 0.0300 &  0.0420 & 0.0530\\
   $\sigma_\mathrm{MAD}$($dz$) & 0.0195 &  0.0266 & 0.0329\\
   $\sigma_\mathrm{NMAD}$($dz$) & 0.0121 &  0.0168 & 0.0211\\
   $\sigma$($d\zeta$) & 0.0300 &  0.0420 & 0.0530\\
   $\sigma_\mathrm{MAD}$($d\zeta$) & 0.0195 &  0.0266 & 0.0329 \\
   $\sigma_\mathrm{NMAD}$($d\zeta$) & 0.0121 &  0.0168 & 0.0211\\

\hline
\end{tabular}
\tablefoot{$dz = (z_\mathrm{phot} - z_\mathrm{spec})/(1+z_\mathrm{spec})$ and $d\zeta = \zeta_\mathrm{phot} - \zeta_\mathrm{spec}$.}
\end{table}

The photo-z performance compared to the spectroscopic sample is usually estimated using three quantities: the median bias, the outlier fraction, and the scatter around the spectroscopic redshift. Photo-z accuracy is generally estimated using the normalised median absolute deviation between the photometric and spectroscopic redshift samples. Mathematically, it is expressed as
\begin{equation}
    \sigma_\mathrm{NMAD} = 1.48 \times \mathrm{median} \left( \left| \Delta \zeta - \mathrm{median}(\Delta \zeta) \right| \right)
    \label{eq:sigma_nmad}
\end{equation}
where $\Delta \zeta$ is the difference between the photometric and spectroscopic redshifts expressed in $\zeta$,
\begin{equation}
\Delta \zeta = \zeta_{\mathrm{phot}} - \zeta_{\mathrm{spec}} .
\end{equation}
The bias in the photo-z estimate is defined as $\mathrm{median}\left(\Delta \zeta\right)$, and outliers are defined where $\left|\Delta \zeta\right|$ exceeds some fixed value. In this paper, we defined the outliers using two thresholds, $\left|\Delta \zeta\right|>0.05$ and $\left|\Delta \zeta\right|>0.1$. The distribution of differences in the GAMA sample is shown in Fig.~\ref{fig:dz_distribution}. In Table~\ref{table:topz_photoz_params} we quantify the photo-z bias, precision, and outlier fraction for three different magnitude bins. The values are given both for redshift $z$ and $\zeta=\ln{\left (1+z\right )}$.

Figure~\ref{fig:outlier_sigmanmad_vs_mag} shows the outlier fraction and $\sigma_\mathrm{NMAD}$ for different magnitude bins. As expected (see Sect.~\ref{sec:photoz_gama}), the $\sigma_\mathrm{NMAD}$ depends on the observed magnitude, while the photo-z accuracy decreases towards fainter magnitudes. The upper panel in Fig.~\ref{fig:outlier_sigmanmad_vs_mag} shows the outlier fraction as a function of galaxy magnitude. For the outlier threshold $\left|\Delta \zeta\right|>0.1$, the fraction of outliers is below two per~cent; the only exception is the faintest magnitude bin, where it is around three per~cent. However, when we look at the outlier fraction defined with $\left|\Delta \zeta\right|>0.05$, the number of outliers significantly increases for galaxies fainter than $m_Z > 17$\,mag. For brighter galaxies, the outlier fraction ($\left|\Delta \zeta\right|>0.05$) remains below four per~cent.

Figure~\ref{fig:zspec_mag_grid} shows the $\sigma_\mathrm{NMAD}$, median bias ($\mathrm{median}\left(\Delta\zeta\right)$) and outlier fraction ($\left|\Delta \zeta\right|>0.1$) as a function of spectroscopic redshift and apparent magnitude. The lowest redshift bin is affected by the photo-z prior $\zeta_\mathrm{phot} > 0.0$ (see Sect.~\ref{sec:photoz_gama} and Fig.~\ref{fig:zspec_vs_zphot}). The faintest magnitude bin and the largest redshift bins contain relatively few sources. We consider these aspects while interpreting the Fig.~\ref{fig:zspec_mag_grid}.

In Fig.~\ref{fig:zspec_mag_grid}, the $\sigma_\mathrm{NMAD}$ and outlier fraction behave as expected, and both of them mainly depend on the apparent magnitude (see also Fig.~\ref{fig:outlier_sigmanmad_vs_mag}). The largest redshift bins have a higher outlier fraction for every magnitude range. However, since the number of sources in those bins is relatively low, we cannot draw any firm conclusions. Looking at how the median bias ($\mathrm{median}\left(\Delta\zeta\right)$) depends on redshift and magnitude, we see that there is a clear dependence on redshift. Additionally, the bias depends on the magnitude for redshifts $\zeta<0.2$. Comparing the bias values with $\sigma_\mathrm{NMAD}$, we see that the bias is always many times smaller than the $\sigma_\mathrm{NMAD}$. Hence, the slight systematic bias does not dominate, and the width of the photo-z posterior predominantly determines the photo-z accuracy. Whether the systematic bias comes from the template-fitting algorithm or the GAMA observational data requires detailed analysis and a large sample of synthetic galaxy spectra.

\begin{figure}
\centering
    \includegraphics[]{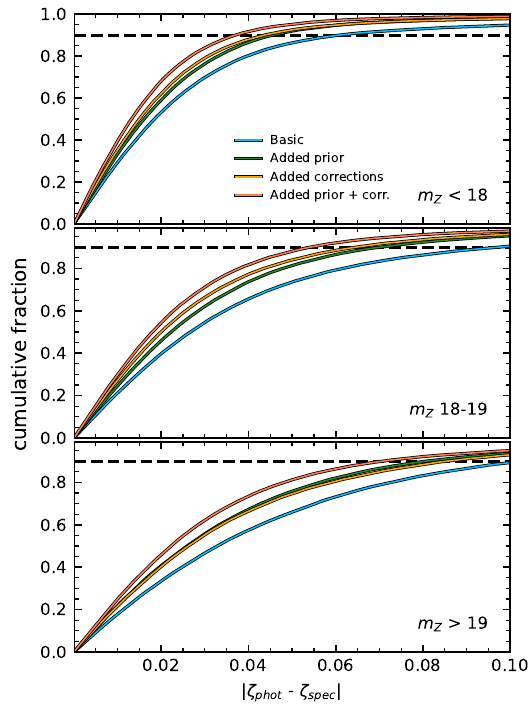}
    \caption{TOPz feature comparison in three magnitude ranges (different panels). The coloured lines show the cumulative fraction of objects with a certain photo-z accuracy. Each coloured line represents a different TOPz run with certain features turned on or off, as described by the legend on the top panel. The three panels have different sets of galaxies based on their Z-band magnitude (as shown in the bottom-right corner of each panel). The dashed black line indicates the 90\% fraction to help better distinguish the accuracies on each panel. See Fig.~\ref{fig:dz_distribution} for the differential distribution of differences between spectroscopic redshifts and photo-z.}
    \label{fig:cumulative_dz_feature_comparison}
\end{figure}

\subsection{Importance of flux corrections and physical priors}
\label{sec:photoz_statistics2}

In Sect.~\ref{sec:fluxcorr}, we described how we correct the input galaxy fluxes and flux uncertainties. In Sect.~\ref{sec:prior}, we introduced a physically motivated prior for the TOPz algorithm. In this section, we will show how these two aspects improve the accuracy of photo-z in general. In our previous paper, \citet{2022A&A...668A...8L} showed that both are improving the photo-z estimation. In \citet{2022A&A...668A...8L}, the method was applied to the miniJPAS data, and the flux corrections and priors were defined differently than in the current paper. Below, we will repeat the same analysis using the GAMA data and use the updated flux corrections and priors as described in this paper.

Figure~\ref{fig:cumulative_dz_feature_comparison} shows the cumulative fraction of objects, where $\left|\Delta\zeta\right|$ is smaller than indicated in the $x$-axis. The different lines show the cumulative fraction for four cases where the flux corrections or priors were turned on or off. It is clear that when both of them are turned off (titled 'basic' on the plot), the photo-z accuracy is significantly lower than when both are used. This is valid for all three magnitude bins. When the flux corrections or priors were turned off individually, the photo-z accuracy was between the two previous cases. We see that both are equally important in improving the photo-z accuracy. For brighter galaxies, the flux corrections have a slightly larger impact.

\begin{figure}
\centering
    \includegraphics[]{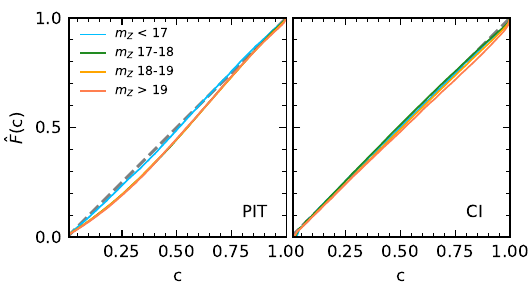}
    \caption{PIT and CI tests that were conducted on the TOPz output PDFs. The two panels from left to right show the cumulative fraction of the underlying PIT and CI values, respectively. For more info on PIT and CI tests, see Sect.~\ref{sect:photoz_pdf_test}. The one-to-one relation that shows the ideal distribution is drawn with a diagonal dashed grey line. The cumulative fraction is given in four magnitude ranges with the values and colours identical to Fig.~\ref{fig:dz_distribution}. Only the faintest magnitude bin shows a slight deviation from the theoretical expectation.}
    \label{fig:pdf_tests}
\end{figure}

\begin{figure}
\centering
    \includegraphics[]{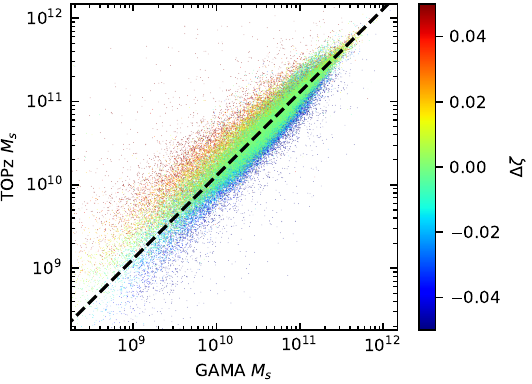}
    \caption{Comparison between TOPz and GAMA stellar masses for each galaxy. The x-axis shows the GAMA stellar masses acquired using GKV ProFound photometry (see Sect.~\ref{sec:gama}). The y-axis shows the TOPz stellar masses that are computed using the spectral templates from the CIGALE software (see Sect.~\ref{sec:phys_params}). The colour denotes the TOPz photo-z accuracy. The dashed black line is a median stellar mass ratio relation, where $M_{\star,\mathrm{TOPz}} \approx 1.3 \cdot M_{\star,\mathrm{GAMA}}$. Galaxies where TOPz photo-z are close to the spectroscopic redshifts follow the stellar masses from the GAMA database very well.}
    \label{fig:stellarmass}
\end{figure}

\begin{figure*}
\centering
    \includegraphics[]{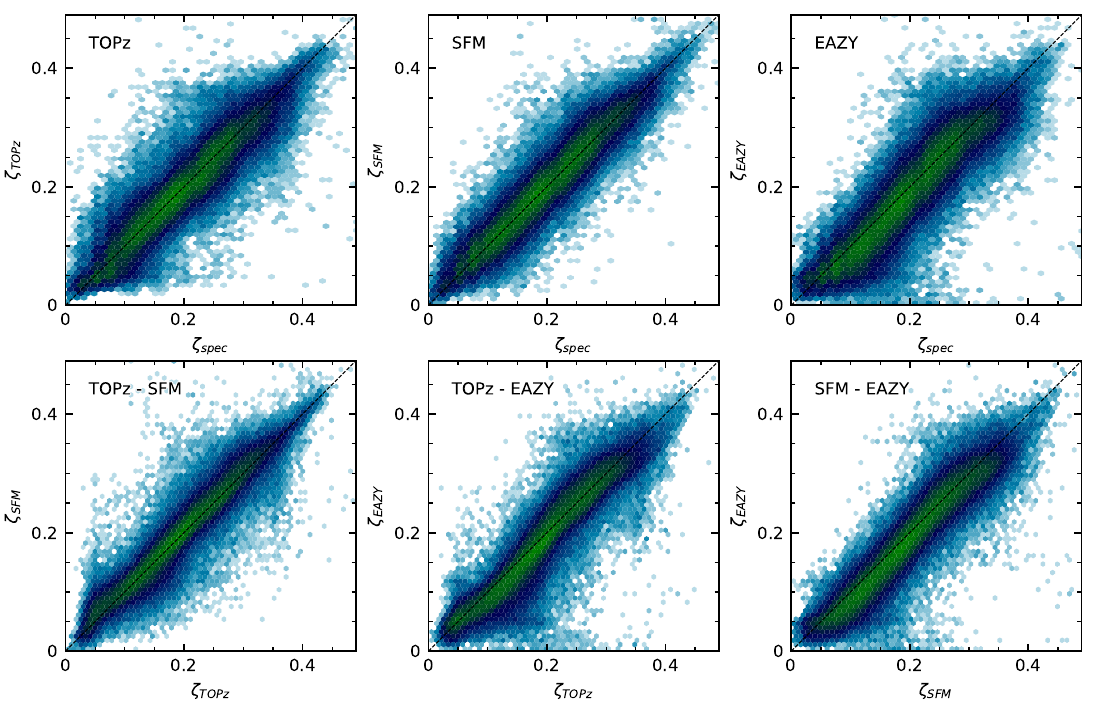}
    \caption{Comparison between photo-z codes using GAMA data. All the colours and markings are identical to the top panels in Fig.~\ref{fig:zspec_vs_zphot}. \textbf{Upper panels}: The relation between photo-z estimates and spectroscopic redshifts for the TOPz, SFM, and EAZY codes, respectively. See Sect.~\ref{sect:gama_eazy} for a brief description of each method. \textbf{Lower panels}: The relation between the photo-z estimates for each photo-z code pair.}
    \label{fig:comp_photoz_codes_hexbins}
\end{figure*}

\subsection{Tests on photometric redshift PDFs}
\label{sect:photoz_pdf_test}

In this section, we analyse how well the photo-z probability distribution functions (posteriors) reflect the actual uncertainty of the estimated photo-z. In Sect.~\ref{sec:photoz_gama} (see Fig.~\ref{fig:dz_distribution} lower panel), we showed that the TOPz photo-z uncertainty computed with Eq.~\eqref{eq:zeta_sigma} adequately estimates the actual difference between the photometric and spectroscopic redshift estimates. However, this was estimated for a single object at a time. In this section, we perform integral tests over the entire sample to assess whether, statistically, the TOPz photo-z posteriors are adequate.

In \citet{2016arXiv160808016P}, the probability integral transform (PIT) is utilised as an analytical tool to assess the calibration and precision of the photo-z PDFs. 
The probability value for any single galaxy $G$ is defined as: 
\begin{equation}
    c_\mathrm{PIT}(G) = \sum\limits_{\zeta<\zeta^G_\mathrm{spec}} p^G_\zeta(\zeta),
    \label{eq:PIT}
\end{equation}
where we assumed that the photo-z PDF is given in a grid and the PDF is normalised to $\sum p_\zeta(\zeta) = 1.0$. The $\zeta_\mathrm{spec}$ is the true redshift, which is the spectroscopic redshift measurement in our case.
A PIT test assesses the distribution of the PIT values for the whole data set.
When plotted as the cumulative fraction of the single galaxy PIT values $\hat{F}(c_\mathrm{PIT})$, a uniform distribution that indicates well-calibrated PDFs will be close to the one-to-one relation. 
However, any significant deviations from that relation indicate that the PDFs are either over- or underconfident.

Another test to evaluate the PDF calibration involves calculating the fraction of galaxies ($\hat{F}(c_\mathrm{CI})$) for which the spectroscopic redshift is within a specified confidence interval (CI).
This approach is detailed in \citet{2016MNRAS.457.4005W}, who define the CI as the area under the PDF that reaches the probability threshold calculated at the spectroscopic redshift.
For a single galaxy $G$, the confidence is defined as:
\begin{equation}
    c_\mathrm{CI}(G) = \sum_{\zeta\in p^G_\zeta(\zeta)\ge p^G_\zeta (\zeta_\mathrm{spec}) } p^G_\zeta (\zeta) .
    \label{eq:CI}
\end{equation}
For well-calibrated PDFs, one would expect 10\% of galaxies to fall within a 10\% CI, 20\% within a 20\% CI, and so on.
Thus, similarly to the PIT test, if the cumulative fraction $\hat{F}(c_\mathrm{CI}$) of the whole data set aligns with the CI values, the PDFs are considered well-calibrated and will be close to the one-to-one relation. 
Again, any deviations below or above this expectation indicate under- or overconfidence, respectively.

\begin{figure}
\centering
    \includegraphics[]{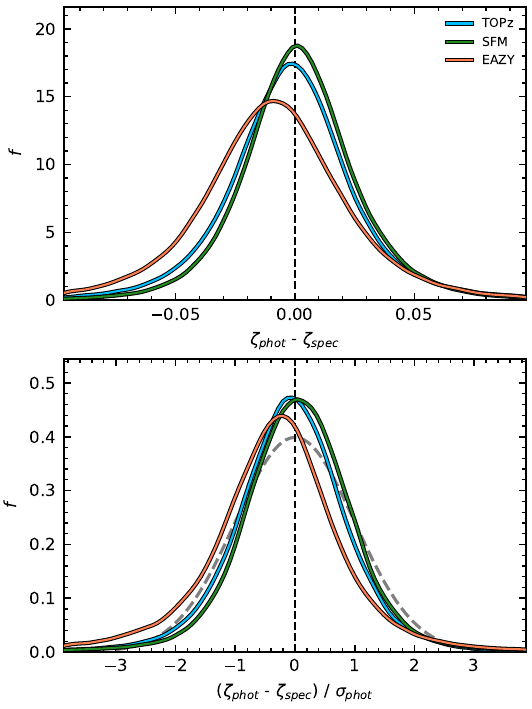}
    \caption{Photo-z accuracy distributions for different photo-z codes. The panels are identical to Fig.~\ref{fig:dz_distribution} except for the colours that now denote the various codes instead.}
    \label{fig:comp_photoz_codes_distribution}
\end{figure}

To assess the calibration of our PDFs, we conducted both the PIT and the CI tests, and the results can be seen in Fig.~\ref{fig:pdf_tests}. We performed the test for four magnitude bins, where only the faintest magnitude bin in the PIT test showed a slight deviation from the one-to-one line. Generally, the photo-z PDFs from the TOPz code are statistically representative and reflect the estimated photo-z's actual uncertainty.

\subsection{Stellar masses from the TOPz code}
\label{sect:gama_stellar_masses}

Template fitting in the TOPz code uses the physical galaxy spectral templates generated by the CIGALE software (see Sect.~\ref{sec:templates}). In Sect.~\ref{sec:phys_params}, we explained how the CIGALE spectral templates can be used to estimate galaxy physical parameters. In this paper, we used the TOPz code to compute the stellar masses for GAMA galaxies. 

In Fig.~\ref{fig:stellarmass}, we compare the TOPz stellar masses with the stellar masses available in the GAMA database (see Sect.~\ref{sec:gama}). The GAMA stellar masses are computed for the spectroscopic sample using the spectroscopic redshifts, while TOPz stellar masses are gathered from the best-fitting templates. In general, TOPz stellar masses follow the GAMA stellar masses very nicely. Most of the scatter visible in Fig.~\ref{fig:stellarmass} is caused by the difference between the photometric and spectroscopic redshifts. The 1.3 times difference between the stellar masses in the GAMA database and from the TOPz code (CIGALE models) is probably caused by the differences between the assumptions, initial mass functions, and the stellar population synthesis models used by the GAMA team and in the CIGALE software.

\subsection{Comparison with EAZY and SFM photometric redshifts}
\label{sect:gama_eazy}

\begin{table}
\caption{Photo-z parameters for the TOPz, EAZY, and SFM codes.}
\label{table:codes_photoz_params}
\centering
\begin{tabular}{l c c c}
\hline\hline
 & TOPz & EAZY & SFM  \\
\hline
\noalign{\vspace{2pt}}
    BIAS  &  & &\\
   median($dz$)  & -0.001609 &  -0.008851 & 0.001390\\
   median($d\zeta$)  & -0.001611 &  -0.008890 & 0.001389\\[4pt]
   OUTLIERS  &  & & \\
   |$dz$| > 0.1  & 1.86\% &  2.69\% & 1.00\%\\
   |$dz$| > 0.15  & 0.71\% &  1.21\%  & 0.37\%\\
   |$d\zeta$| > 0.1  & 1.83\% &  2.80\% & 0.97\% \\
   |$d\zeta$| > 0.15  & 0.68\% &  1.24\% & 0.35\% \\
   |$d\zeta$|$ / \sigma_\mathrm{phot}$ > 3  & 2.06\% &  5.79\% & 3.04\%\\[4pt]
   PRECISION  &  & & \\
   $\sigma$($dz$) & 0.0370 &  0.1440 & 0.0310\\
   $\sigma_\mathrm{MAD}$($dz$) & 0.0241 &  0.0305 & 0.0221\\
   $\sigma_\mathrm{NMAD}$($dz$) & 0.0152 &  0.0188 & 0.0139\\
   $\sigma$($d\zeta$) & 0.0370 &  0.0810 & 0.0310\\
   $\sigma_\mathrm{MAD}$($d\zeta$) & 0.0241 &  0.0306 & 0.0221 \\
   $\sigma_\mathrm{NMAD}$($d\zeta$) & 0.0152 &  0.0190 & 0.0139\\

\hline
\end{tabular}
\tablefoot{$dz = (z_\mathrm{phot} - z_\mathrm{spec})/(1+z_\mathrm{spec})$ and $d\zeta = \zeta_\mathrm{phot} - \zeta_\mathrm{spec}$.}
\end{table}

The photo-z for the GAMA data has been previously computed using the EAZY code and the Scaled Flux Matching (SFM) technique. See \citet{2022MNRAS.513..439D} for a detailed description of the GAMA data release. We will use these two previous photo-z estimates and compare them with the photometric redshifts derived using the TOPz code. EAZY and TOPz codes are both based on the template-fitting method. The main difference between the two codes is in their details. Both codes use different spectral templates, flux corrections and priors\footnote{The EAZY code applied to the GAMA sample includes no flux calibration offsets (E.~N.~Taylor, private communication).}. The second method, the SFM technique, relies on the spectroscopic sample and assigns the photo-z based on the similarity of the observed galaxy SED in different passbands. In general, the SFM technique is closer to the traditional machine-learning photo-z codes, where the main limitation is the coverage of the spectroscopic sample. For the GAMA sample used in the current paper, the spectroscopic coverage is excellent, and it is expected that the SFM code performs better than the template-fitting approach.

To compare the three codes, we use the spectroscopic sample where we have the photo-z estimate from all three codes, which limits our sample to 188\,984 galaxies. From EAZY code, we used the $z_\mathrm{peak}$ values and from SFM code, the $\mathrm{zexp}$ values. The performance of the TOPz, SFM, and EAZY codes can be seen in the upper panel of Fig.~\ref{fig:comp_photoz_codes_hexbins}. In general, all three codes are comparable, while the TOPz seems to slightly outperform the EAZY results. Looking at the outliers, it is visible that the SFM is performing the best, which is somewhat expected since the spectroscopic training sample for the SFM technique is sufficiently representative of the GAMA spectroscopic sample. The lower panel in Fig.~\ref{fig:comp_photoz_codes_hexbins} shows the comparison between the three codes themselves. It is visible that the distribution is narrower than for the comparison with the spectroscopic sample. This indicates that the photometry limits the photo-z accuracy in all codes. Since all codes use the same photometry, they are expected to provide very similar photo-z estimates. A more detailed look reveals that the two codes most identical to each other are the TOPz and SFM. In Table~\ref{table:codes_photoz_params} we quantify the bias, precision, and outlier fraction for the three codes.

The photo-z accuracy for the TOPz, SFM, and EAZY codes is shown in Fig.~\ref{fig:comp_photoz_codes_distribution}. The distribution of differences between photometric and spectroscopic redshifts is shown in the upper panel, where one can see that the EAZY code has the largest bias and the distribution is wider than for the other two codes. The TOPz and SFM distributions are centred around zero and have fairly similar widths, while SFM shows marginally better results. On the lower panel of Fig.~\ref{fig:comp_photoz_codes_distribution}, we normalised the redshift difference with photo-z uncertainty. For SFM code, we used the $\mathrm{zerr}$ values as the $\sigma_\mathrm{phot}$. For EAZY code, we calculated the $\sigma_\mathrm{phot}$ as $(\mathrm{u}68 - \mathrm{l}68) / 2$. For both cases, we took care to convert the $\sigma_\mathrm{phot}$ values from the redshift units to $\zeta$ units. The normalised distributions confirm that the TOPz and SFM codes perform equally well. The normalised distributions are slightly more peaked than the Gaussian distribution, indicating that the photo-z uncertainty might be slightly underestimated for some sources.

The TOPz performs very well, outperforming slightly the similar template-fitting code EAZY. Compared with SFM, the two codes are equally good, while SFM has a slightly lower number of outliers. It is somewhat expected since SFM is similar to the machine-learning codes and is trained using the spectroscopic redshift sample. A clear path forward to improve the TOPz photo-z is to combine the template-fitting methods with an approach that uses the spectroscopic redshift sample.

In the analysis carried out in this section, the TOPz used the GAMA spectroscopic sample to optimise the galaxy template library. Similarly, SFM used the same spectroscopic sample for photo-z estimation. When the same spectroscopic dataset is used for comparison, both codes are expected to perform very well. However, the photo-z estimates are not guaranteed to be reliable for fainter galaxies and at higher redshifts. In Appendix~\ref{app:desi} we use the independent DESI spectroscopic redshifts to validate the TOPz photo-z estimates for fainter and slightly higher redshift galaxies. As shown in the Appendix~\ref{app:desi}, the conclusions drawn in this section are valid.

\section{Conclusions}
\label{sec:conclusions}

In this paper, we have introduced the updated TOPz code, a Bayesian photo-z estimation tool based on template fitting. Our approach integrates several critical features that significantly enhance the accuracy and reliability of photo-z estimates.

We have demonstrated the necessity of both flux corrections and flux uncertainty corrections to account for systematic biases and uncertainties in the observed photometric measurements. These corrections ensure that the input photometry is accurate and unbiased for the SED fitting. Additionally, the physical prior, derived from the luminosity function and survey volume prior, improves the redshift estimates by incorporating expected galaxy distributions. This approach is crucial for achieving reliable photo-z, especially for faint galaxies with more than one peak in their photo-z posterior.

A vital aspect of any template-fitting photo-z codes is the selection of spectral templates. We have demonstrated the importance of template set optimisation in adequately representing the diversity of galaxy types. This optimisation eliminates the need for an independent template prior and ensures that the templates cover all possible galaxy types (see Section~\ref{sec:templates}). The physical templates generated using the CIGALE software provide a good basis for the template set optimisation, allowing us to cover a wide variety of galaxy spectral types needed for template-fitting photo-z estimation.

Our results show that the TOPz photo-z posteriors are statistically reliable and accurately reflect the uncertainty in photo-z estimates. The photo-z accuracy across various magnitude bins shows that the flux corrections and posteriors effectively capture the actual uncertainty of the estimates (see Section~\ref{sec:topz_gama}).

Applying TOPz to the GAMA spectroscopic sample, we found no significant bias in the photo-z estimates (see Table~\ref{table:topz_photoz_params}). The $\sigma_\mathrm{NMAD}$ values are uniform, and the outlier fraction is low, indicating the robustness of our approach. The $\sigma_\mathrm{NMAD}=0.012$ for $m_Z<18$ and increases to $\sigma_\mathrm{NMAD}=0.021$ for $m_Z>19$. The outlier fraction ($|dz|>0.1$) in the same magnitude bins increases from 1\% to 5\%. Table~\ref{table:topz_photoz_params} lists more quantitative measures. The comparison with other photo-z codes, such as EAZY and SFM, shows that TOPz performs competitively, providing accurate and reliable redshift estimates. In our comparison with the EAZY and SFM photo-z codes, we found that TOPz performs very well, slightly outperforming the EAZY code and showing comparable results to the SFM technique (see Table~\ref{table:codes_photoz_params} for a quantitative comparison). The latter, which relies on the spectroscopic sample for training, had a slightly lower number of outliers, which is expected given its similarity to machine-learning approaches. However, the TOPz and SFM codes both show excellent performance, with the main differences attributed to the underlying methodologies and the use of spectroscopic training samples.

Using the physical templates from CIGALE, we estimated the stellar masses for GAMA galaxies. The comparison with the GAMA database shows that TOPz stellar masses are consistent, with most of the scatter being attributed to differences between photometric and spectroscopic redshifts. The ability to estimate physical parameters, such as stellar masses, directly from photometric data highlights the potential of the TOPz approach.

In this paper, we used the GAMA sample as a proof-of-concept. Extending this methodology to other surveys requires additional testing and validation. To further improve photo-z accuracy, we plan to incorporate additional photometric data, address systematics in photometry, and reduce outliers. Including machine-learning methodology in the template-fitting code is also a promising avenue. We aim to apply the TOPz algorithm to the J-PAS narrow-band survey and explore the combination of the J-PAS narrow bands with SDSS photometry. This will validate and enhance the capabilities of the TOPz code, particularly for high-redshift galaxies. In Appendix~\ref{app:desi} we show an independent comparison between TOPz photo-z and DESI DR1 spectroscopic redshifts, which extend to higher redshifts and fainter magnitudes not used in the TOPz photo-z optimisation process.

In summary, the TOPz code significantly advances photo-z estimation based on the template-fitting approach. By integrating flux corrections, physical priors, and optimised template sets, TOPz provides accurate, reliable, and physically motivated redshift estimates. These capabilities are crucial for advancing our understanding of the universe through cosmological and galaxy evolution studies. Combining the spectroscopic redshift samples with photometric redshift data allows us to study the faint dwarf galaxy population in the cosmic web. This is a promising avenue in the upcoming 4MOST WAVES and 4HS surveys \citep{2019Msngr.175....3D, 2019Msngr.175...46D, 2023Msngr.190...46T}.

\section*{Data availability}

Table presented in Appendix~\ref{app:gama} is only available in electronic form at the CDS via anonymous ftp to cdsarc.u-strasbg.fr (130.79.128.5) or via \url{http://cdsweb.u-strasbg.fr/cgi-bin/qcat?J/A+A/}.

The TOPz code is entirely written in modern Fortran and is publicly available on GitHub (\url{https://github.com/etempel/TOPz}). In addition to the code, we make available all Python scripts that were used to generate the figures presented in this paper. The figure scripts are part of the same GitHub repository. All the internal data sets used during the paper preparation are available upon reasonable request. The filter transmission curves used in this paper are taken from the Spanish Virtual Observatory Filter Profile Service \citep{2012ivoa.rept.1015R, 2020sea..confE.182R}.

\begin{acknowledgements}
The authors thank the anonymous referee for their detailed reading of the paper and helpful recommendations, resulting in multiple improvements. We thank Edward (Ned) Taylor for valuable discussions about applying the EAZY code to the GAMA sample. This work was funded by the Estonian Ministry of Education and Research (grant TK202), Estonian Research Council (grants PRG1006, PRG3034, PRG2159, PSG700) and the European Union's Horizon Europe research and innovation programme (EXCOSM, grant No. 101159513). Extensive exploratory analysis in this work was done using the TOPCAT software \citep{2005ASPC..347...29T}. Most of the figures were generated using the Matplotlib package \citep{Hunter:2007}.
GAMA is a joint European-Australasian project based around a spectroscopic campaign using the Anglo-Australian Telescope. The GAMA input catalogue is based on data from the Sloan Digital Sky Survey and the UKIRT Infrared Deep Sky Survey. Complementary imaging of the GAMA regions is being obtained by a number of independent survey programmes, including GALEX MIS, VST KiDS, VISTA VIKING, WISE, Herschel-ATLAS, GMRT and ASKAP, providing UV to radio coverage. GAMA is funded by the STFC (UK), the ARC (Australia), the AAO, and the participating institutions. The GAMA website is \href{https://www.gama-survey.org}{https://www.gama-survey.org}.
This research has made use of the Spanish Virtual Observatory (\href{https://svo.cab.inta-csic.es}{https://svo.cab.inta-csic.es}) project funded by MCIN/AEI/10.13039/501100011033 through grant PID2020-112949GB-I00.
\end{acknowledgements}

\bibliographystyle{aa}
\bibliography{topz_references}

\begin{appendix}

\section{TOPz photo-z for the GAMA sample}
\label{app:gama}

The TOPz photo-z catalogue for the GAMA photometric sample is made available through the Strasbourg Astronomical Data Centre (CDS). The output table contains the following information (column numbers are given in square brackets):
\begin{enumerate}
 \item{[1]\,\texttt{uberID} --} unique GAMA\,III ID of object;
 \item{[2]\,\texttt{CATAID} --} ID of best matching GAMA\,II object;
 \item{[3]\,\texttt{zspec} --} spectroscopic redshift from the GAMA database;
 \item{[4--5]\,\texttt{ra, dec} --} right ascension and declination (deg);
 \item{[6]\,\texttt{mag} --} magnitude in the $r+Z$ detection band;
 \item{[7]\,\texttt{nflux} --} number of used fluxes for photo-z estimation;
 \item{[8]\,\texttt{zeta\_w} --} posterior-weighted redshift ($\zeta=\ln{(1+z)}$);
 \item{[9]\,\texttt{mass\_w} --} posterior-weighted stellar mass estimate;
 \item{[10]\,\texttt{zeta\_best} --} posterior-weighted redshift from the best peak;
 \item{[11]\,\texttt{zeta\_best\_p} --} summed probability in the best peak;
 \item{[12--13]\,\texttt{zeta\_best\_min, zeta\_best\_max} --} minimum and maximum best-peak redshift values;
 \item{[14]\,\texttt{zeta\_best\_sigma} --} weighted standard deviation of the best-peak redshift;
 \item{[15]\,\texttt{zeta\_alt1} --} posterior-weighted redshift from the first alternative peak;
 \item{[16]\,\texttt{zeta\_alt1\_p} --} summed probability in the first alternative peak;
 \item{[17--18]\,\texttt{zeta\_alt1\_min, zeta\_alt1\_max} --} minimum and maximum first alternative peak redshift values;
 \item{[19]\,\texttt{zeta\_alt1\_sigma} --} weighted standard deviation of the first alternative peak;
 \item{[20]\,\texttt{zeta\_alt2} --} posterior-weighted redshift from the second alternative peak;
 \item{[21]\,\texttt{zeta\_alt2\_p} --} summed probability in the second alternative peak;
 \item{[22--23]\,\texttt{zeta\_alt2\_min, zeta\_alt2\_max} --} minimum and maximum second alternative peak redshift values;
 \item{[24]\,\texttt{zeta\_alt2\_sigma} --} weighted standard deviation of the second alternative peak redshift;
 \item{[25]\,\texttt{npeak} --} number of identified peaks in the photo-z posterior;
 \item{[26]\,\texttt{chi2\_mean} --} posterior-weighted mean $\chi^2$ value;
 \item{[27]\,\texttt{is\_star} --} TOPz probability that this object is a star.
\end{enumerate}

\section{TOPz and EAZY results in the DESI DR1 sample}
\label{app:desi}

\begin{figure*}
\centering
    \includegraphics[]{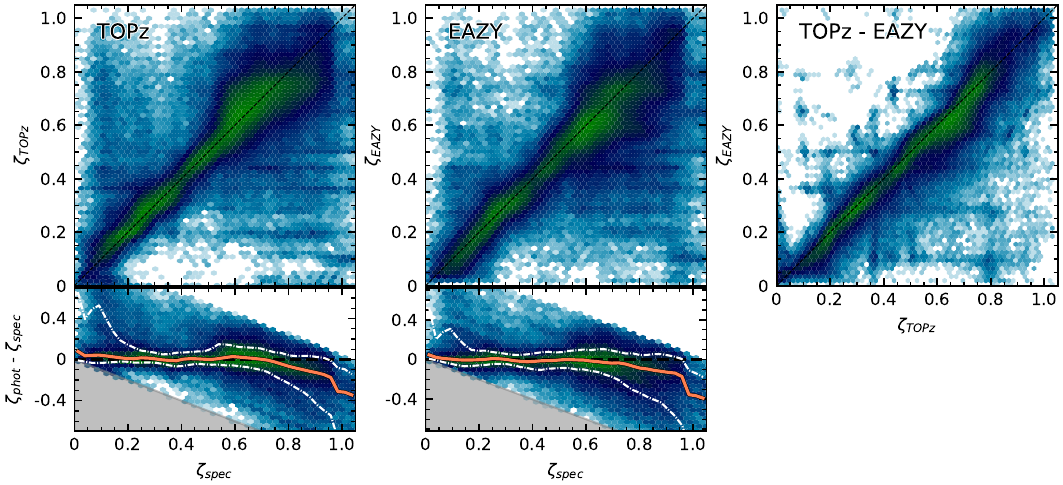}
    \caption{Comparison between photo-z results of TOPz and EAZY in the DESI sample. \textbf{Upper panels}: The two leftmost panels show the individual photo-z results compared to the spectroscopic redshift ($\zeta=\ln(1+z)$) while the right panel shows the relation between the TOPz and EAZY photo-z. The layout is similar to Fig.~\ref{fig:comp_photoz_codes_hexbins}. \textbf{Lower panels}: Photo-z accuracy as a function of spectroscopic redshift for TOPz and EAZY codes, respectively. The layout is similar to the lower panels in Fig.~\ref{fig:zspec_vs_zphot}.}
    \label{fig:desi_eazy_topz_photoz}
\end{figure*}

\begin{figure*}
\centering
    \includegraphics{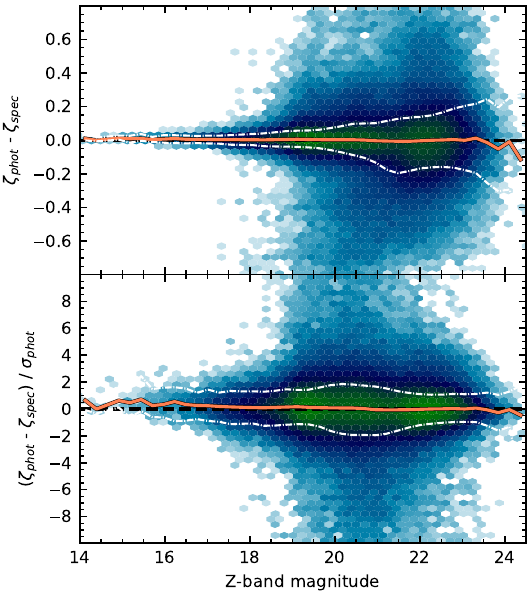}
    \includegraphics{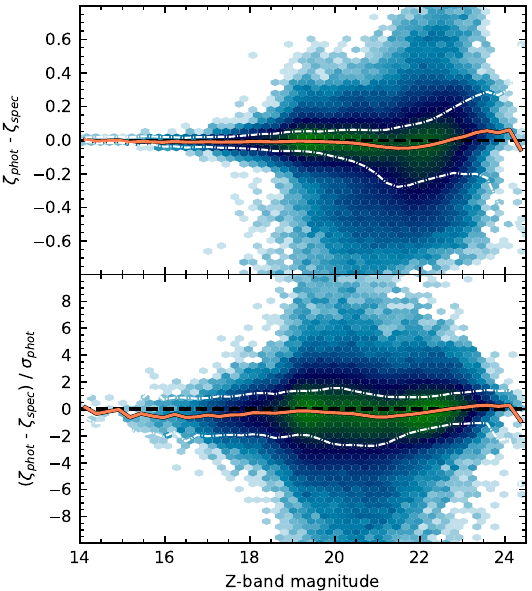}
    \caption{Magnitude dependence on the photo-z accuracy for TOPz (left) and EAZY (right) code in the DESI sample. The layout is identical to Fig.~\ref{fig:dz_vs_mag}.}
    \label{fig:DESI_mag_dependene_topz}
\end{figure*}

\begin{figure}
\centering
    \includegraphics[]{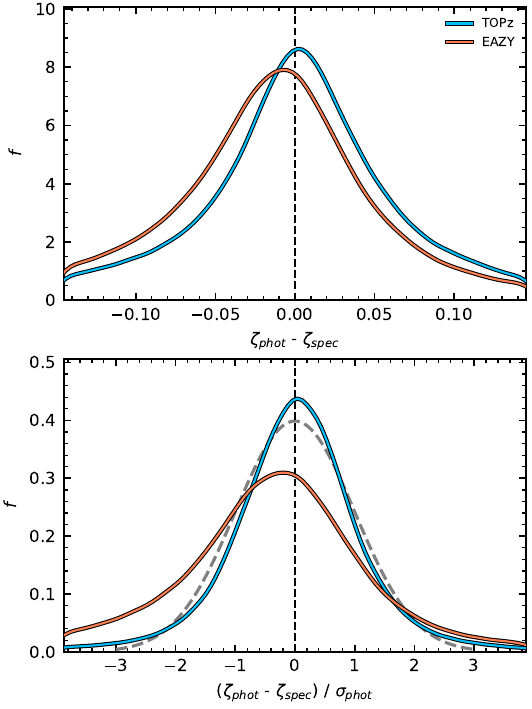}
    \caption{Photo-z accuracy for the TOPz and EAZY codes in the DESI sample. The layout is identical to Fig.~\ref{fig:comp_photoz_codes_distribution}.}
    \label{fig:desi_eazy_topz_distribution}
\end{figure}

To validate the TOPz and EAZY photo-z results, we use the DESI DR1 (Dark Energy Spectroscopic Instrument Data Release~1)\footnote{\url{https://data.desi.lbl.gov/doc/releases/dr1/}} \citep{2025arXiv250314745D}. We select DESI sources, where $\mathrm{ZCAT\_PRIMARY}=True$, $\mathrm{ZWARN}=0$ and $\mathrm{OBJTYPE}=\mathrm{TGT}$. We limited our sample to $z<2.0$. To have an independent validation set, we eliminated sources that were part of the GAMA spectroscopic sample.

In Figs.~\ref{fig:desi_eazy_topz_photoz} and \ref{fig:DESI_mag_dependene_topz} we show the comparison between DESI, EAZY, and TOPz redshift estimates. Visually, the scatter between TOPz and EAZY is smaller than between photo-z and DESI spectroscopic redshift. In Fig.~\ref{fig:desi_eazy_topz_distribution}, we show the absolute and uncertainty-scaled difference as a function of magnitude. While both EAZY and TOPz perform similarly, there is a slightly smaller bias in TOPz photo-z estimates. This is also confirmed quantitatively in Table~\ref{table:desi_photoz_params}. In Fig.~\ref{fig:desi_eazy_topz_distribution}, we show the distribution of uncertainty-scaled differences between EAZY and TOPz photo-z and DESI spectroscopic redshifts. It is visible that TOPz results are less biased and the uncertainties follow more closely the expected Gaussian distribution.

The analysis carried out using independent DESI spectroscopic redshifts confirms that the conclusions drawn in the main body of the paper are valid. It is true even for the fainter and higher redshift galaxies that were not used during the TOPz photo-z optimisation process.

\begin{table}
\caption{Photo-z parameters for DESI sample.}
\label{table:desi_photoz_params}
\centering
\begin{tabular}{l c c}
\hline\hline
 & TOPz & EAZY  \\
\hline
\noalign{\vspace{2pt}}
    BIAS  &  & \\
   median($dz$)  & 0.000357 &  -0.018056  \\
   median($d\zeta$)  & 0.000357 &  -0.018221  \\[4pt]
   OUTLIERS  &  &  \\
   |$dz$| > 0.1  & 24.97\% &  27.07\%  \\
   |$dz$| > 0.15  & 15.71\% &  17.64\%  \\
   |$d\zeta$| > 0.1  & 24.94\% &  27.50\%  \\
   |$d\zeta$| > 0.15  & 15.77\% &  18.20\%  \\
   |$d\zeta$|$ / \sigma_\mathrm{phot}$ > 3  & 7.62\% &  19.74\%  \\[4pt]
   PRECISION  &  &  \\
   $\sigma$($dz$) & 0.156 &  0.149  \\
   $\sigma_\mathrm{MAD}$($dz$) & 0.0644 &  0.0712  \\
   $\sigma_\mathrm{NMAD}$($dz$) & 0.0467 &  0.0506  \\
   $\sigma$($d\zeta$) & 0.174 &  0.180  \\
   $\sigma_\mathrm{MAD}$($d\zeta$) & 0.0644 &  0.0716  \\
   $\sigma_\mathrm{NMAD}$($d\zeta$) & 0.0467 &  0.0511  \\

\hline
\end{tabular}
\tablefoot{$dz = (z_\mathrm{phot} - z_\mathrm{spec})/(1+z_\mathrm{spec})$ and $d\zeta = \zeta_\mathrm{phot} - \zeta_\mathrm{spec}$.}
\end{table}

\end{appendix}

\end{document}